\documentclass[a4paper,10pt,
twocolumn,colorlinks,urlcolor=black,linkcolor=black,citecolor=black,
]{article}
\usepackage{amsmath,amssymb}
\usepackage[dvips]{color}
\usepackage{cite}
\usepackage{hangcaption}
\usepackage{amsmath,amssymb}
\usepackage[dvips]{color}
\usepackage{cite}
\usepackage{eepic}
\usepackage{epic}
\usepackage{hangcaption}
\usepackage{amsmath}
\usepackage{amssymb}
\usepackage{amscd}
\usepackage{amsthm}
\usepackage[dvipdfmx,%
 bookmarks=true,%
 bookmarksnumbered=true,%
 colorlinks=true,%
 setpagesize=false,%
 pdfkeywords={TeX; dvipdfmx; hyperref; color;}]{hyperref}
\def\roman{\rm}

\def\cases{\left\{\begin{array}{ll}}
\def\endcases{\end{array}\right.}
\def\bigtimes{\mathop{\mbox{\Large $\times$}}}
\begin{document}
\twocolumn[
\vskip0.5cm
\centerline{\LARGE \bf The Linguistic Interpretation of Quantum Mechanics}
\vskip0.5cm
\begin{center}
{\rm
\large
Shiro Ishikawa
}
\\
\vskip0.2cm
\rm
\it
Department of Mathematics, Faculty of Science and Technology,
Keio University, 
\\
3-14-1 Hiyoshi, Kohoku-ku, Yokohama, 223-8522 Japan ({E-Mail:
ishikawa@math.keio.ac.jp})
\\
\end{center}
\par
\rm
\vskip0.3cm
\par
\noindent
{}
\normalsize
\par
$\;\;\;$About twenty years ago,
we proposed the mathematical formulation of Heisenberg's uncertainty principle,
and further, we concluded that Heisenberg's uncertainty principle
and EPR-paradox are not contradictory. This is true, however we now think that 
we should have argued about it under a certain firm interpretation of quantum mechanics. Recently we proposed the linguistic quantum interpretation
(called quantum and classical measurement theory), which was characterized as a kind of metaphysical and linguistic turn of the Copenhagen interpretation.
This turn from physics to language does not only extend quantum theory to classical systems but also yield the quantum mechanical world view
(i.e., the philosophy of quantum mechanics, in other words,  
quantum philosophy). In fact, we can consider that
traditional philosophies have progressed toward quantum philosophy.
In this paper, we first review the linguistic quantum interpretation, 
and further, clarify the relation between EPR-paradox and Heisenberg's uncertainty principle. That is, the linguistic interpretation says that
EPR-paradox is closely related to the fact that
syllogism does not generally hold in quantum physics.
This fact should be compared to the non-locality of Bell's inequality.
\par
\vskip0.3cm
\par
\noindent
\par
\vskip1.0cm
\par
]

\par
\vskip0.3cm
\par
\noindent
{
\begin{itemize}
\item[{\large \bf 1.}]
{
\large
\bf
Introduction
}
\end{itemize}
}
\par
About twenty years ago
( in 1991),
we proposed and proved the mathematical formulation
of Heisenberg's uncertainty principle
({\it cf.}
Theorem 2 and Corollary 1 in ref. {\cite{Ishi1}}).
As mentioned in Section 4.6
(i.e.,
inequalities
\hyperref[eq8]{{{{}}}{(8)}}
and
\hyperref[eq9]{{{{}}}{(9)}}
)
later,
note
that
Heisenberg's uncertainty principle
should not be 
\par
\noindent
\begin{picture}(500,170)
\thicklines
\put(0,70){
$
\fbox{{\shortstack[l]{world \\ 
view}}}
$
$
\left\{\begin{array}{l}
\!\!\!
\textcircled{\scriptsize R}:
\underset{\text{\scriptsize (realism)}}{\fbox{\text{Aristotle}}}
{\xrightarrow[]{\textcircled{\scriptsize 1}}}
\overset{\text{\scriptsize (monism)}}{\underset{\text{\scriptsize (realism)}}
{\fbox{\text{Newton}}}}
\xrightarrow[]{}
\left\{\begin{array}{llll}
\fbox{\shortstack[l]{theory of \\ relativity}}
\xrightarrow[]{\qquad \qquad \quad \qquad \qquad}
\\
\\
{\fbox{\shortstack[l]{quantum \\ mechanics}}}
{
\xrightarrow[]{}
\left\{\begin{array}{lll}
\xrightarrow[\quad \text{\scriptsize \rm realistic view}\;\;]{{\text{\rm }}}
\\
\\
\xrightarrow[{{\textcircled{\scriptsize 3}}\rm :\text{linguistic turn}}]{
{\textcircled{\scriptsize 2}}\rm:
{dualism}}
\end{array}\right.
}
\end{array}\right.
\\
\\
\!\!\!
\textcircled{\scriptsize L}:
\overset{\text{\scriptsize }}{
\underset{\text{\scriptsize (idealism)}}{\fbox{\shortstack[l]{Plato \\ 
Parmenides}}}
}
\xrightarrow[]{{\textcircled{\scriptsize 4}}}
\overset{\text{\scriptsize (dualism)}}{
\underset{\text{\scriptsize (idealism)}}{\fbox{
{\shortstack[l]{Kant \\ Descartes}}
}}
}
{\xrightarrow[]{\textcircled{\scriptsize 5}}}
\underset{\text{\scriptsize (linguistic view)}}{\fbox{
\shortstack[l]{philosophy \\ of language}
}}
\xrightarrow[{{\textcircled{\scriptsize 7}}:\text{linguistic view}}]{{
{\textcircled{\scriptsize 6}}:
\rm
axiomatization}}
\end{array}\right.
$
}
\put(332,110){
$
\left.\begin{array}{llll}
\; 
\\
\; 
\\
\; 
\\
\;
\end{array}\right\}
\xrightarrow[]{}
\overset{\text{\scriptsize (unsolved)}}{
\underset{\text{\scriptsize (physics)}}{
\fbox{\shortstack[l]{theory of \\ everything}}
}
}
$
}
\put(332,40){
$
\left.\begin{array}{lllll}
\; 
\\
\; 
\\
\; 
\\
\;
\\
\;
\end{array}\right\}
{\xrightarrow[]{\textcircled{\scriptsize 8}}}
\underset{\text{\scriptsize (scientific language)}}{
\fbox{\shortstack[l]{measurement \\ theory (=MT)}}
}
$
}
\put(10,-20){
{\bf \hypertarget{fig1}{Figure 1}. 
\rm
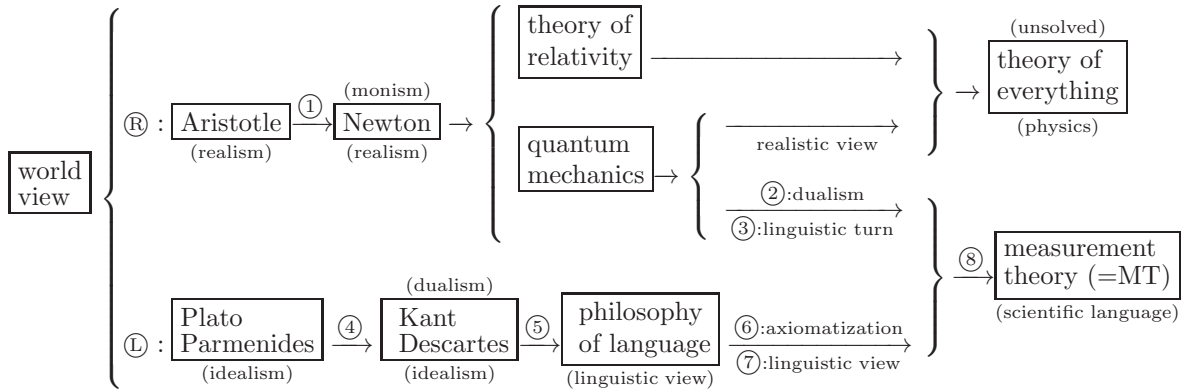
Traditional philosophies from the quantum philosophical point of view
(cf. Section 5)}
}
\end{picture}
\par
\newpage
\par
\noindent
confused with
Robertson's uncertainty principle
{{{}}}{\cite{Robe}}.
And further,
in the remark 3 
of {{{}}}{\cite{Ishi1}},
we discussed
the relation between
Heisenberg's uncertainty principle
and EPR-paradox {{{}}}{\cite{Eins}},
and
concluded that the two are not contradictory.
This is true, but now we consider that
the argument 
in the remark 3 of {{{}}}{\cite{Ishi1}}
is somewhat shallow.
In order to
clarify
the relation
between
Heisenberg's uncertainty principle
and

\newpage
\par
\noindent
EPR-paradox,
we must start from the proposal of the interpretation
of quantum mechanics.
That is 
because
even the Copenhagen
interpretation is not determined
uniquely.
This is our motivation to propose
the linguistic interpretation of quantum mechanics
{{{}}}{\cite{Ishi2,Ishi3}}.

In this paper,
adding new examples,
we first review the linguistic interpretation
{{{}}}{\cite{Ishi2,Ishi3}}, 
and explain \hyperlink{fig1}{\bf Figure 1}.
And lastly,
we
clarify the relation between
EPR-paradox and Heisenberg's uncertainty 
principle
in the linguistic interpretation of quantum mechanics.
\vskip0.2cm
\par
\noindent
{
\begin{itemize}
\item[{\large \bf 2.}]
{
\large
\bf
Measurement Theory
(Axioms
)
}
\end{itemize}
}
\par
\par
\noindent
\begin{itemize}
\item[{\bf 2.1.}]
\bf
Overview: Measurement theory
\end{itemize}

\rm
\par
\par
\noindent
In this section,
we shall explain the overview of measurement theory
(or in short, MT).
\par
\par
\rm
Measurement theory
(refs.{{{}}}{\cite{Ishi2,Ishi3,Ishi4,Ishi5,Ishi6, Ishi7, Ishi8, Ishi9}})
is,
by an analogy of
quantum mechanics
(or,
as a linguistic turn of quantum
mechanics
), constructed
as the scientific
theory
formulated
in a certain 
{}{$C^*$}-algebra ${\cal A}$
(i.e.,
a norm closed subalgebra
in the operator algebra $B(H)$
composed of all bounded operators on a Hilbert space $H$,
{\it cf.$\;$}{{{}}}{\cite{Murp, Neum}}
).
MT is composed of
two theories
(i.e.,
pure measurement theory
(or, in short, PMT]
and
statistical measurement theory
(or, in short, SMT).
That is,
we see:
\par
\rm
\par
\begin{itemize}
\item[{{{}}}{\hypertarget{A}{(A)}}]
$
\;\;
\underset{\text{\footnotesize }}{
\text{
MT (measurement theory)
}
}
$
\\
\\
\rm
$=\underset{\text{\rm \scriptsize (Axiom 1)}}{\text{[measurement]}}
+
\underset{\text{\scriptsize (Axiom 2)}}{\text{[causality]}}$
\\
\\
$=\cases
\text{(A$_1$)}:
\underset{\text{\scriptsize }}{\text{[PMT
]}}
\\
=
\displaystyle{
{
\mathop{\mbox{[(pure) measurement]}}_{\text{\scriptsize (Axiom$^{\rm P}$ 1) }}
}
}
+
\displaystyle{
\mathop{
\mbox{
[causality]
}
}_{
{
\mbox{
\scriptsize
(Axiom 2)
}
}
}
}
\\
\\
\text{(A$_2$)}
:
\underset{\text{\scriptsize }}{\text{[SMT
]}}
\\
=
\displaystyle{
{
\mathop{\mbox{[(statistical) measurement]}}_{\text
{\scriptsize (Axiom$^{\rm S}$ 1) }}
}
}
\!
+
\!
\displaystyle{
\mathop{
\mbox{
[causality]
}
}_{
{
\mbox{
\scriptsize
(Axiom 2)
}
}
}
}
\endcases
$
\end{itemize}
where
Axiom 2 is common in PMT and SMT.
For completeness, note that measurement theory {{{}}}{(\hyperlink{A}{A})}
(i.e.,
{{{}}}{(\hyperlink{A}{A$_1$})} and {{{}}}{(\hyperlink{A}{A$_2$})})
is
a kind of language
based on
{\lq\lq}the quantum mechanical world view{\rq\rq}.
It may be understandable
to
consider that
\begin{itemize}
\item[{{{}}}{\hypertarget{B1}{(B$_1$)}}]
PMT and SMT
is related to
Fisher's statistics
and
Bayesian statistics
respectively.
\end{itemize}
This is discussed in \cite{Ishi4}.
Thus, if we believe in \hyperlink{fig1}{\bf Figure 1},
we can answer to the following problem
({\it cf.} \cite{Ishi4}):
\begin{itemize}
\item[{{{}}}{\hypertarget{B2}{(B$_2$)}}]
What is statistics?
$\;\;$
Or,
where is statistics in science?
\end{itemize}

%

\par
\noindent
\begin{itemize}
\item[{\bf 2.2.}]
\bf
Mathematical preparations
\end{itemize}

In this paper,
we mainly devote ourselves to
the $C^*$-algebraic formulation,
and not the $W^*$-algebraic formulation
({\it cf.} the appendix in \cite{Ishi2}).
Thus, Axiom$^{\text P}$ 1 is often denoted by Axiom 1.

When ${\cal A}=B_c(H)$,
the ${C^*}$-algebra composed
of all compact operators on a Hilbert space $H$,
the {{{}}}{\hyperlink{A}{(A)}} is called {quantum measurement theory}
(or,
quantum system theory),
which can be regarded as
the linguistic aspect of quantum mechanics.
Also, when ${\cal A}$ is commutative
$\big($
that is, 
when ${\cal A}$ is characterized by $C_0(\Omega)$,
the $C^*$-algebra composed of all continuous 
complex-valued functions vanishing at infinity
on a locally compact Hausdorff space $\Omega$
({\it cf.$\;$}{{{}}}{\cite{Murp}})$\big)$,
the {{{}}}{\hyperlink{A}{(A)}} is called {classical measurement theory}.
Thus, we have the following classification:
\begin{itemize}
\item[{{{}}}{\hypertarget{}{(C)}}]
$
\quad
\underset{\text{\scriptsize }}{\text{MT}}
$
$\left\{\begin{array}{ll}
\text{quantum MT$\quad$(when ${\cal A}=B_c (H)$)}
\\
\\
\text{classical MT
$\quad$
(when ${\cal A}=C_0(\Omega)$)}
\end{array}\right.
$
\end{itemize}
In this paper, we mainly devote ourselves to classical MT
(i.e.,
classical PMT and classical SMT).

\par
\noindent
\par
Now we shall explain the measurement theory
{{{}}}{\hyperlink{A}{(A)}}.
Let
${\cal A}
( \subseteq B(H))$
be
a
${C^*}$-algebra,
and let
${\cal A}^*$ be the
dual Banach space of
${\cal A}$.
That is,
$ {\cal A }^* $
$ {=}  $
$ \{ \rho \; | \; \rho$
is a continuous linear functional on ${\cal A}$
$\}$,
and
the norm $\| \rho \|_{ {\cal A }^* } $
is defined by
$ \sup \{ | \rho ({}F{}) |  \:{}: \; F \in {\cal A}
\text{ such that }\| F \|_{{\cal A}} 
(=\| F \|_{B(H)} )\le 1 \}$.
The bi-linear functional
$\rho(F)$
is
also denoted by
${}_{{\cal A}^*}
\langle \rho, F \rangle_{\cal A}$,
or in short
$
\langle \rho, F \rangle$.
Define the
\it
mixed state
$\rho \;(\in{\cal A}^*)$
\rm
such that
$\| \rho \|_{{\cal A}^* } =1$
and
$
\rho ({}F) \ge 0
\text{ 
for all }F\in {\cal A}
\text{ satisfying }
F \ge 0$.
And put
\begin{align} {\frak S}^m  ({}{\cal A}^*{})
{=}
\{ \rho \in {\cal A}^*  \; | \;
\rho
\text{ is a mixed state}
\}.
\label{eq1}
\end{align}
%
\rm
A mixed state
$\rho (\in {\frak S}^m  ({\cal A}^*) $)
is called a
\it
pure state
\rm
if
it satisfies that
{\lq\lq $\rho = \theta \rho_1 + ({}1 - \theta{}) \rho_2$
for some
$ \rho_1 , \rho_2 \in {\frak S}^m  ({\cal A}^*)$
and
$0 < \theta < 1 $\rq\rq}
implies
{\lq\lq $\rho =  \rho_1 = \rho_2$\rq\rq}\!.
Put
\begin{align} {\frak S}^p  ({}{\cal A}^*{})
{=}
\{ \rho \in {\frak S}^m  ({\cal A}^*)  \; | \;
\rho
\text{ is a pure state}
\},
\label{eq2}
\end{align}
which is called a
\it
state space.
\rm
\rm
The Riesz theorem ({\it cf.}{{{}}}{\cite{Yosi}}) says that
$C_0(\Omega )^*$
$=$
${\cal M}(\Omega )
$
$=$
$\{
\rho \;|\;
\rho
$
is a signed measure on $\Omega$
$
\}$,
${\frak S}^m(C_0(\Omega )^*)$
$=$
${\cal M}_{+1}^m(\Omega )
$
$=$
$\{
\rho \;|\;
\rho
$
is a measure on $\Omega$
such that
$\rho(\Omega)=1$
$
\}$.
Also,
it is well known
({\it cf.}{{{}}}{\cite{Murp}})
that
$ {\frak S}^p  ({}{B_c(H)}^*{})=$
$\{ | u \rangle \! \langle u |
$
(i.e., the Dirac notation)
$
\:\;|\;\:
$
$
\|u \|_H=1 
\}$,
and
$ {\frak S}^p  ({}{C_0(\Omega)}^*{})$
$={\cal M}_{+1}^p(\Omega)$
$=$
$\{ \delta_{\omega_0} \;|\; \delta_{\omega_0}$ is a point measure at
${\omega_0}
\in \Omega
\}$,
where
$ 
\int_\Omega f(\omega) \delta_{\omega_0} (d \omega )$
$=$
$f({\omega_0})$
$
(\forall f
$
$
\in C_0(\Omega))$.
The latter implies that
$ {\frak S}^p  ({}{C_0(\Omega)}^*{})$
can be also identified with
$\Omega$
(called a {\it spectrum space}
or simply
{\it spectrum})
such as
\begin{align}
\underset{\text{\scriptsize (state space)}}{{\frak S}^p  ({}{C_0(\Omega)}^*{})}
\ni \delta_{\omega} \leftrightarrow {\omega} \in 
\underset{\text{\scriptsize (spectrum)}}{\Omega}
\label{eq3}
\end{align}

Here, assume
that
the
${C^*}$-algebra
${\cal A}
( \subseteq B(H))$
is unital,
i.e.,
it
has the identity $I$.
This assumption is not unnatural,
since, if $I \notin {\cal A}$,
it suffices to reconstruct the ${\cal A}$ such that it includes 
${\cal A}\cup \{I\}$.
In this sense,
the $C_0(\Omega)$ is often denoted by
$C(\Omega)$.

According to the noted idea ({\it cf.}{{{}}}{\cite{ Davi}})
in quantum mechanics,
an {\it observable}
${\mathsf O}{\; \equiv}(X, {\cal F},$
$F)$ in 
${{\cal A}}$
is defined as follows:
\par
\par
\begin{itemize}
\item[{{{}}}{\hypertarget{D1}{(D$_1$)}}]
[Field]
$X$ is a set,
${\cal F}
(\subseteq 2^X $,
the power set of $X$)
is a field of $X$,
that is,
{\lq\lq}$\Xi_1, \Xi_2 \in {\cal F}\Rightarrow \Xi_1 \cup \Xi_2 \in {\cal F}${\rq\rq},
{\lq\lq}$\Xi  \in {\cal F}\Rightarrow X \setminus \Xi \in {\cal F}${\rq\rq}.
\item[{{{}}}{\hypertarget{D2}{(D$_2$)}}]
[Countably additivity]
$F$ is a mapping from ${\cal F}$ to ${{\cal A}}$ 
satisfying:
(a):
for every $\Xi \in {\cal F}$, $F(\Xi)$ is a non-negative element in 
${{\cal A}}$
such that $0 \le F(\Xi) $
$\le I$, 
(b):
$F(\emptyset) = 0$ and 
$F(X) = I$,
where
$0$ and $I$ is the $0$-element and the identity
in ${\cal A}$
respectively.
(c):
for any countable decomposition $\{ \Xi_1,\Xi_2, \ldots \}$
of $\Xi$
$\in {\cal F}$
(i.e., $\Xi_k , \Xi \in {\cal F}$
such that
$\bigcup_{k=1}^\infty \Xi_k = \Xi$,
$\Xi_i \cap \Xi_j= \emptyset
(i \not= j)$),
it holds that
\end{itemize}
\begin{align}
&
\quad
\lim_{K \to \infty } \rho( F( \bigcup_{k=1}^K \Xi_k ))
=
\rho( F( \Xi ))
\quad
(
\forall \rho \in {\frak S}^m  ({\cal A}^*)
)
\nonumber
\\
&
\quad
\;\;\;
\text{
(i.e., in the sense of weak convergence).
}
\label{eq4}
\end{align}
\par
\noindent
\par
\vskip0.3cm
\par
\par
\noindent
{\it Remark 1.}
By the Hopf extension theorem
({\it cf.$\;$}{{{}}}{\cite{Yosi}}),
we have the mathematical probability space
$(X,$
$ {\overline{\cal F}},$
$ \rho^m (F(\cdot )) \;)$ 
where
${\overline{\cal F}}$
is the smallest $\sigma$-field such that
${{\cal F}} \subseteq {\overline{\cal F}}$.
For the other formulation
(i.e.,
$W^*$-algebraic formulation
),
see, for example,
the appendix in {{{}}}{\cite{Ishi2}}.

\par
\noindent
\begin{itemize}
\item[{\bf 2.3.}]
\bf
Pure measurement theory
in {{{}}}{(\hyperlink{A}{A$_1$})}
\end{itemize}
\rm
\par
In what follows,
we shall explain PMT
in
{{{}}}{(\hyperlink{A}{A$_1$})}.
\par
\rm
With any {\it system} $S$, a $C^*$-algebra 
${\cal A}( \subseteq B(H))$ can be associated in which the 
pure
measurement theory {{{}}}{(\hyperlink{A}{A$_1$})} of that system can be formulated.
A {\it state} of the system $S$ is represented by an element
$\rho (\in {\frak S}^p  ({}{\cal A}^*{}))$
and an {\it observable} is represented by an observable 
${\mathsf{O}}{\; =} (X, {\cal F}, F)$ in ${{\cal A}}$.
Also, the {\it measurement of the observable ${\mathsf{O}}$ for the system 
$S$ with the state $\rho$}
is denoted by 
${\mathsf{M}}_{{{\cal A}}} ({\mathsf{O}}, S_{[\rho]})$
$\big($
or more precisely,
${\mathsf{M}}_{\cal A} ({\mathsf{O}}{\; :=} (X, {\cal F}, F), S_{[\rho]})$
$\big)$.
An observer can obtain a measured value $x $
($\in X$) by the measurement 
${\mathsf{M}}_{\cal A} ({\mathsf{O}}, S_{[\rho]})$.
\par
\noindent
\par
The Axiom$^{\rm P}$ 1 presented below is 
a kind of mathematical generalization of Born's probabilistic interpretation of quantum mechanics.
And thus, it is a statement without reality.
\par
\noindent
{\bf{Axiom$^{\rm P}$ 1\;\;
\rm
$[$Pure Measurement$]$}}.
\it
The probability that a measured value $x$
$( \in X)$ obtained by the measurement 
${\mathsf{M}}_{{{\cal A}}} ({\mathsf{O}}$
${ :=} (X, {\cal F}, F),$
{}{$ S_{[\rho_0]})$}
belongs to a set 
$\Xi (\in {\cal F})$ is given by
$
\rho_0( F(\Xi) )
$.
\rm

\par
\par
\vskip0.2cm
\par

\par
Next, we explain Axiom 2 in {{{}}}{\hyperlink{A}{(A)}}.
Let $(T,\le)$ be a tree, i.e., a partial ordered 
set such that {\lq\lq$t_1 \le t_3$ and $t_2 \le t_3$\rq\rq} implies {\lq\lq$t_1 \le t_2$ or $t_2 \le t_1$\rq\rq}\!.
In this paper,
we assume that
$T$ is finite.
Assume that
there exists an element $t_0 \in T$,
called the {\it root} of $T$,
such that
$t_0 \le t$ ($\forall t \in T$) holds.
Put $T^2_\le = \{ (t_1,t_2) \in T^2{}\;|\; t_1 \le t_2 \}$.
The family
$\{ \Phi_{t_1,t_2}{}: $
${\cal A}_{t_2} \to {\cal A}_{t_1} \}_{(t_1,t_2) \in T^2_\le}$
is called a {\it causal relation}
({\it due to the Heisenberg picture}),
\rm
if it satisfies the following conditions 
{{{}}}{(\hyperlink{E1}{E$_1$})} and 
{{{}}}{(\hyperlink{E2}{E$_2$})}.
\begin{itemize}
\rm
\item[{{{}}}{\hypertarget{E1}{(E$_1$)}}]
With each
$t \in T$,
a $C^*$-algebra ${\cal A}_t$
is associated.
\item[{{{}}}{\hypertarget{E2}{(E$_1$)}}]
For every $(t_1,t_2) \in T_{\le}^2$, a Markov operator 
$\Phi_{t_1,t_2}{}: {\cal A}_{t_2} \to {\cal A}_{t_1}$ 
is defined
(i.e.,
$\Phi_{t_1,t_2} \ge 0$,
$\Phi_{t_1,t_2}(I_{{\cal A}_{t_2}})$
$
=
$
$
I_{{\cal A}_{t_1}}$
).
And it satisfies that
$\Phi_{t_1,t_2} \Phi_{t_2,t_3} = \Phi_{t_1,t_3}$ 
holds for any $(t_1,t_2)$, $(t_2,t_3) \in T_\le^2$.
\end{itemize}
\noindent
The family of dual operators
$\{ \Phi_{t_1,t_2}^*{}: $
$
{\frak S}^m  ({\cal A}_{t_1}^*)
\to {\frak S}^m  ({\cal A}_{t_2}^*)
\}_{(t_1,t_2) \in T^2_\le}$
is called a
{
\it
dual causal relation}
({\it
due to the Schr\"{o}dinger picture}).
When
$ \Phi_{t_1,t_2}^*{}$
$
(
{\frak S}^p  ({\cal A}_{t_1}^*)
)$
$\subseteq
$
$
(
{\frak S}^p  ({\cal A}_{t_2}^*)
)$
holds for any
$
{(t_1,t_2) \in T^2_\le}$,
the causal relation is said to be
deterministic.

\par
\par
\rm
Now Axiom 2 in the measurement theory {{{}}}{\hyperlink{A}{(A)}} is presented
as follows:
\rm
\par
\noindent
{\bf{Axiom 2}
\rm[Causality]}.
\it
The causality is represented by
a causal relation 
$\{ \Phi_{t_1,t_2}{}: $
${\cal A}_{t_2} \to {\cal A}_{t_1} \}_{(t_1,t_2) \in T^2_\le}$.

\rm
\par

\par
\noindent
\par
\noindent
\par
\noindent
\vskip0.2cm
\par

\noindent
\begin{itemize}
\item[{\bf 2.4.}]
\bf
Statistical measurement theory
in {{{}}}{(\hyperlink{A}{A$_2$})}
\end{itemize}

\rm


We shall introduce the following notation:
\rm
It is usual to consider that
we do not know the pure state
$\rho_0^p$
$(
\in
{\frak S}^p  ({}{\cal A}^*{})
)$
when
we take a measurement
${\mathsf{M}}_{{{\cal A}}} ({\mathsf{O}}, S_{[\rho_0^p]})$.
That is because
we usually take a measurement ${\mathsf{M}}_{{{\cal A}}} ({\mathsf{O}},
S_{[\rho_0^p]})$
in order to know the state $\rho_0^p$.
Thus,
when we want to emphasize that
we do not know the state $\rho_0^p$,
${\mathsf{M}}_{{{\cal A}}} ({\mathsf{O}}, S_{[\rho_0^p]})$
is denoted by
${\mathsf{M}}_{{{\cal A}}} ({\mathsf{O}}, S_{[\ast]})$.
Also,
when we know the distribution $\rho_0^m$
$( \in {\frak S}^m({\cal A}^*) )$
of the unknown state
$\rho_0^p$,
the
${\mathsf{M}}_{{{\cal A}}} ({\mathsf{O}}, S_{[\rho_0^p]})$
is denoted by
${\mathsf{M}}_{{{\cal A}}} ({\mathsf{O}}, S_{[\ast]}
(\{ \rho_0^m \}) )$.
The $\rho_0^m$
and
${\mathsf{M}}_{{{\cal A}}} ({\mathsf{O}}, S_{[\ast]}
(\{ \rho_0^m \}) )$
is respectively called a mixed state
and
a statistical measurement.

\par
\vskip0.3cm

\par
The Axiom$^{\rm S}$ 1 presented below is 
a kind of mathematical generalization of 
Axiom$^{\rm P}$ 1.

\par

\par
\noindent
{\bf{Axiom$^{\rm S}$\;1\;
\rm
\;[Statistical measurement]}}.
\it
The probability that a measured value $x$
$( \in X)$ obtained by the statistical measurement 
${\mathsf{M}}_{{{\cal A}}} ({\mathsf{O}}$
${ \equiv} (X, {\cal F}, F),$
{}{$ S_{[\ast]}(\{ \rho_0^m \}) )$}
%
belongs to a set 
$\Xi (\in {\cal F})$ is given by
$
\rho_0^m ( F(\Xi) )
$
$($
$=
{}_{{{\cal A}^*}}\langle
\rho_0^m,
F(\Xi)
\rangle_{{\cal A}}$
$)$.
\rm


\par

Here,
Axiom 2 and Interpretation 
{{{}}}{\hyperlink{H}{(H)}} mentioned in the following section are common in {{{}}}{(\hyperlink{A}{A$_1$})} and
{{{}}}{(\hyperlink{A}{A$_2$})}.

%

\rm
\par
\noindent

\par
\par
\vskip0.5cm
\par
\par
\noindent
{\bf
\large
3. The Linguistic Interpretation
\\
$\quad$
(\hyperlink{fig1}{\bf Figure 1}:\textcircled{\scriptsize 2},\textcircled{\scriptsize 3},\textcircled{\scriptsize 6})
}

\par
\noindent
\vskip0.2cm
\par
According to
\cite{Ishi2},
we shall explain the linguistic interpretation of quantum mechanics.
The measurement theory {{{}}}{(\hyperlink{A}{A})} asserts
\begin{itemize}
\item[{{{}}}{\hypertarget{F}{(F)}}]
Obey
Axioms 1 and 2.
And,
describe any ordinary phenomenon according to
Axioms 1 and 2
(in spite that
Axioms 1 and 2 can not be tested experimentally).
\end{itemize}
Still,
most readers 
may be perplexed how to use Axioms 1 and 2
since there are various usages.
Thus, the following problem is significant.
\begin{itemize}
\item[{{{}}}{\hypertarget{G}{(G)}}]
How should Axioms 1 and 2
be used?
\end{itemize}
Note that reality is not reliable
%
since Axioms 1 and 2 are statements without reality.

Here, in spite of the linguistic turn
(\hyperlink{fig1}{\bf Figure 1}:\textcircled{\scriptsize 3})
and the mathematical generalization
from
$B(H)$ to
a $C^*$-algebra
${\cal A}$,
we consider that
the dualism (i.e.,
the spirit of so called Copenhagen interpretation)
of quantum mechanics
is
inherited
to measurement theory
(\hyperlink{fig1}{\bf Figure 1}:\textcircled{\scriptsize 2},
or also, see
{{{}}}{\hyperlink{I}{(I)}} later
).
Thus, we present the following interpretation
{{{}}}{\hypertarget{H}{(H)}}
[={{{}}}{{{{}}}{(\hyperlink{H1}{H$_1$})}--{{{}}}{(\hyperlink{H3}{H$_3$})}}].
That is,
as the answer to the question
{{{}}}{\hyperlink{G}{(G)}}, we propose:
\begin{itemize}
\item[{{{}}}{\hypertarget{H1}{(H$_1$)}}]
Consider the dualism composed of {\lq\lq}observer{\rq\rq} and {\lq\lq}system( =measuring object){\rq\rq}.
And therefore,
{\lq\lq}observer{\rq\rq} and {\lq\lq}system{\rq\rq}
must be absolutely separated.
\item[{{{}}}{\hypertarget{H2}{(H$_2$)}}]
Only one measurement is permitted.
And thus,
the state after a measurement
is meaningless
$\;$
since it 
can not be measured any longer.
Also, the causality should be assumed only in the side of system,
however,
a state never moves.
Thus,
the Heisenberg picture should be adopted.
\item[{{{}}}{\hypertarget{H3}{(H$_3$)}}]
Also, the observer
does not have
the space-time.
Thus, 
the question:
{\lq\lq}When and where is a measured value obtained?{\rq\rq}
is out of measurement theory,
\end{itemize}
\par
\noindent
and so on.

Although
N. Bohr (i.e.,
the chief proponent of
the Copenhagen interpretation
)
said,
in the Bohr--Einstein debates
{{{}}}{\cite{Bohr, Eins}},
that
the interpretation of a physical theory has to rely on 
an experimental practice.
However,
we consider that
all confusion is due to
the preconception
that
the Copenhagen interpretation is within physics.
In this sense,
we agree with A. Einstein,
who never accepted the Copenhagen interpretation as
physics.
That is, in spite of
Bohr's realistic view,
we propose the following linguistic world view
(\hyperlink{fig1}{\bf Figure 1}:\textcircled{\scriptsize 3})
:
\begin{itemize}
\item[{{{}}}{\hypertarget{I}{(I)}}]
In the beginning was the language called measurement theory
(with the interpretation {{{}}}{\hyperlink{H}{(H)}}).
And, for example, quantum mechanics can be
fortunately
described in this language.
And moreover,
almost all scientists have already mastered this language
partially and informally
since statistics
(at least, its basic part)
is characterized as one of aspects of
measurement theory
({\it cf.} 
{{{}}}{\cite{Ishi2,Ishi3, Ishi4,Ishi5,Ishi6, Ishi7,Ishi8,
Ishi9 }}
).
\end{itemize}
In this sense,
we consider that
measurement theory holds as a kind of
language-game
(with the rule
(Axioms 1 and 2, Interpretation {{{}}}{\hyperlink{H}{(H)}}),
and therefore,
measurement theory
is regarded as
the axiomatization (\hyperlink{fig1}{\bf Figure 1}:\textcircled{\scriptsize 6}) of
the philosophy of language
(i.e., Saussure's linguistic world view).

\par
\vskip1.0cm
\par
\par
\noindent
{\bf
\large
4. How to Use the Linguistic Interpretation {{{}}}{\hyperlink{H}{(H)}}}
\par

\par
\noindent
\vskip0.2cm
\par
\noindent
{\bf
4.1.
Parallel measurement, the law of large numbers
}
\par
\par
\noindent
\vskip0.2cm
\par
For each
$k=1,2,...,K$,
consider a measurement
${\mathsf{M}}_{{{{\cal A}_k}}} ({\mathsf{O}_k}{\; :=} (X_k, {\cal F}_k, F_k),$
{{{}}}{$ S_{[\rho_k]}$}).
However,
the interpretation {{{}}}{(\hyperlink{H2}{H$_2$})}
says that
only one measurement is permitted.
Thus,
we consider the spatial tensor $C^*$-algebra
${\text{\large $\otimes$}}_{k=1}^K 
{{{\cal A}_k}}
(\subseteq B(
{\text{\large $\otimes$}}_{k=1}^K H_k))
$,
and 
%
consider the product space
${\text{\large $\times$}}_{k=1}^K X_k$
and the product field
$\text{\large $\boxtimes$}_{k=1}^K {\cal F}_k$,
which is defined by
the smallest field that contains
a family
$\{ {\text{\large $\times$}}_{k=1}^K {\Xi}_k
\;|
\;
\Xi_k \in {\cal F}_k, k=1,2,...,K\}$.
Define the parallel observable
$\text{\large $\otimes$}_{k=1}^K {\mathsf{O}_k}$
$=({\text{\large $\times$}}_{k=1}^K X_k, $
$\text{\large $\boxtimes$}_{k=1}^K {\cal F}_k, {\widetilde F})$
in the tensor $C^*$-algebra
$\text{\large $\otimes$}_{k=1}^K
{{{\cal A}_k}}
$ 
such that
\begin{eqnarray*}
{\widetilde F}({\text{\large $\times$}}_{k=1}^K {\Xi}_k)
=
\text{\large $\otimes$}_{k=1}^K F_k(\Xi_k)
\\
\;\;
(\forall 
\Xi_k \in {\cal F}_k
,
k=1,2,...,K
).
\end{eqnarray*}
Then, the above
$\{{\mathsf{M}}_{{{{\cal A}_k}}} ({\mathsf{O}_k},$
$ S_{[\rho_k]})$
$\}_{k=1}^K$
is represented 
\par
\noindent
by the parallel measurement${\mathsf{M}}_{{{
\text{$\otimes$}_{k=1}^K 
{{{\cal A}_k}}
}}} (\text{\large $\otimes$}_{k=1}^K{\mathsf{O}_k}
{{{}}}{,}
$
$
S_{[\text{$\otimes$}_{k=1}^K \rho_k]})$,
which is also denoted by
$\text{$\otimes$}_{k=1}^K$
${\mathsf{M}}_{{{{\cal A}_k}}} ({\mathsf{O}_k},$
$ S_{[\rho_k]})$.
Consider a particular case
such that,
${\cal A}={\cal A}_k$,
${\mathsf O}=(X, {\cal F}, F)=(X_k, {\cal F}_k, F_k)$,
$\rho=\rho_k$
($\forall k=1,2,...,K$).
Let $(x_1, x_2, ...,x_K) (\in X^K )$
be a measured value by the 
parallel measurement
$\text{$\otimes$}_{k=1}^K$
${\mathsf{M}}_{{{{\cal A}}}} ({\mathsf{O}},$
$ S_{[\rho]})$.
Then,
using Axiom 1,
we see the law of large numbers, that is,
for sufficiently large $K$,
$$
\rho( F(\Xi)) \approx \frac{
\sharp[\{ 
k=1,2,...,K \;|\;
x_k \in \Xi \}]}{K}
\quad
(\forall \Xi \in {\cal F})
$$
holds,
where
$\sharp[A]$ is the the number of elements of the set $A$.
This is, of course, most fundamental in science.
Also, 
this is the reason that the term {\lq\lq}probability" is 
used in Axiom 1.

\par
\noindent
\vskip0.1cm
\par
\noindent
{\bf
4.2.
Maximal likelihood estimation
}
\par
\par
\rm
\par
\noindent
\par

Consider the classical cases
(i.e.,
${\cal A}=C_0(\Omega)$
).
It may be usual to consider that
Axiom 1 leads 
the following statement,
%
i.e.,
maximum likelihood estimation in classical measurements:
\begin{itemize}
\item[{{{}}}{\hypertarget{J}{(J)}}]
[Maximum likelihood estimation in classical PMT];
When we know that
a measured value obtained by a measurement
${\mathsf{M}}_{{{C_0(\Omega)}}} ({\mathsf{O}}_1$
${\; :=}(X_1, {\cal F}_1, {F}_1),S_{[\ast]})$
belongs to
$\Xi_1 (\in {\cal F}_1)$,
there is a reason to infer that
the unknown state 
[$\ast$]=$\delta_{\omega_0} (\in 
{\frak S}^p  ({}{C_0(\Omega)}^*{})
\approx \Omega,$
by \hyperref[eq3]{(3)}$)$
where
$\omega_0 (\in 
\Omega
)$
is defined by
$
[{F}_1(\Xi_1 )](\omega_0)
=
\max_{
\omega \in \Omega}
[{F}_1(\Xi_1) ](\omega)
$
if it exists.
\end{itemize}
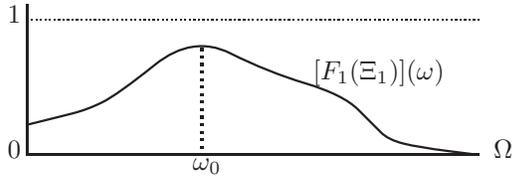
\unitlength=0.20mm
\begin{picture}(400,130)
\put(27,18){0}
\put(27,108){1}
\put(350,18){$\Omega$}
\dottedline{3}(40,110)(340,110)
\put(150,10){$\omega_0$}
\put(40,20){\line(0,1){100}}
\thicklines
\put(40,20){\line(1,0){300}}
\thicklines
\spline(40,40)(60,45)(80,50)(100,60)
(150,100)(200,75)(250,60)
(270,40)(280,30)(300,25)(340,20)
\dottedline{5}(156,20)(156,90)
\put(230,70){$[{}F_1(\Xi_1){}]({}\omega{})$}
\end{picture}
\par
\noindent
Although this {{{}}}{\hyperlink{J}{(J)}} is surely handy,
note that
the {{{}}}{(\hyperlink{H2}{H$_2$})} says that
it is illegal to
regard 
the $\rho_0$ as the state after the measurement ${\mathsf{M}}_{{{C_0(\Omega)}}} ({\mathsf{O}}_1,
S_{[\ast]})$.
Thus,
strictly speaking,
the {{{}}}{\hyperlink{J}{(J)}} is informal.

\rm
\par
\par
\par
\noindent
\par
By a similar method as the lead of the {{{}}}{\hyperlink{J}{(J)}},
we can easily see the following statement
{{{}}}{\hyperlink{K}{(K)}}, which should be regarded
as the measurement 
theoretical form of
maximum likelihood estimation
({\it cf.} {{{}}}{\cite{Ishi4}}, or Corollary 5.5 in {{{}}}{\cite{Ishi8}}).
\begin{itemize}
\item[{{{}}}{\hypertarget{K}{(K)}}]
[Maximum likelihood estimation in general PMT];
When we know that
a measured value obtained by a measurement
${\mathsf{M}}_{{{\cal A}}} ($
${\mathsf{O}}{\; :=}(X_1 {\text{\large $\times$}} X_2,
{\cal F}_1 \boxtimes {\cal F}_2, {F}),$
$ S_{[\ast]})$
belongs to
$\Xi_1 \times X_2$,
there is a reason to infer that
\par
\noindent
the probability that
the measured value 
belongs to
$\Xi_1 \times \Xi_2$
$(\forall \Xi_2 \in {\cal F}_2)$
is given by
the following conditional
probability:
\begin{align}
\frac{\rho_0 ( {F}(\Xi_1 \times \Xi_2))}{
\rho_0 ( {F}(\Xi_1 \times X_2))
}
\label{eq5}
\end{align}
where
$\rho_0 (\in {\frak S}^p  ({}{\cal A}^*{}) )$
is defined by
$
\rho_0 ({F}(\Xi_1 \times X_2)) = 
\max_{\rho \in {\frak S}^p  ({}{\cal A}^*{})}
\rho ( {F}(\Xi_1 \times X_2) )
$
if it exists.
Here, note that
the $\rho_0$
is not
the state after
the measurement
${\mathsf{M}}_{{{\cal A}}} ($
${\mathsf{O}},$
$ S_{[\ast]})$.
\end{itemize}
This {{{}}}{\hyperlink{K}{(K)}}, which also includes quantum cases, is
most fundamental in statistics,
and thus,
we believe
({\it cf.} {{{}}}{\cite{Ishi4, Ishi7}})
that statistics is one of aspects of
measurement theory.
If it be so,
we can answer to the problem {{{}}}{\hyperlink{B2}{(B$_2$)}}.
For the relation between the informal {{{}}}{\hyperlink{J}{(J)}} and the formal {{{}}}{\hyperlink{K}{(K)}}, see Remark 2 later.

\noindent
\begin{itemize}
\item[{\bf 4.3.}]
\bf
Simultaneous measurement
\end{itemize}

\par
For each
$k=1,$
$2,\ldots,K$,
consider a measurement
${\mathsf{M}}_{{{\cal A}}} ({\mathsf{O}_k}$
${\; :=} (X_k, {\cal F}_k, F_k),$
$ S_{[\rho]})$.
However,
since
the {{{}}}{(\hyperlink{H2}{H$_2$})}
says that
only one measurement is permitted,
the
measurements
$\{
{\mathsf{M}}_{{{\cal A}}} ({\mathsf{O}_k},S_{[\rho]})
\}_{k=1}^K$
should be reconsidered in what follows.
Under the commutativity condition such that
\begin{align}
&
F_i(\Xi_i) F_j(\Xi_j) 
=
F_j(\Xi_j) F_i(\Xi_i)
\label{eq6}
%
\\
&
\quad
(\forall \Xi_i \in {\cal F}_i,
\forall \Xi_j \in  {\cal F}_j , i \not= j),
\nonumber
\end{align}
we can
define the product observable
${\text{\large $\times$}}_{k=1}^K {\mathsf{O}_k}$
$=({\text{\large $\times$}}_{k=1}^K X_k ,$
$ \boxtimes_{k=1}^K {\cal F}_k,$
$ 
{\text{\large $\times$}}_{k=1}^K {F}_k)$
in ${\cal A}$ such that
\begin{align*}
({\text{\large $\times$}}_{k=1}^K {F}_k)({\text{\large $\times$}}_{k=1}^K {\Xi}_k )
=
F_1(\Xi_1) F_2(\Xi_2) \cdots F_K(\Xi_K)
\\
\;
(
\forall \Xi_k \in {\cal F}_k,
\forall k=1,\ldots,K
).
\qquad
\qquad
\nonumber
\end{align*}
Here,
$ \boxtimes_{k=1}^K {\cal F}_k$
is the smallest field including
the family
$\{
{\text{\large $\times$}}_{k=1}^K \Xi_k
$
$:$
$\Xi_k \in {\cal F}_k \; k=1,2,\ldots, K \}$.
Then, 
the above
$\{
{\mathsf{M}}_{{{\cal A}}} ({\mathsf{O}_k},S_{[\rho]})
\}_{k=1}^K$
is,
under the commutativity condition \hyperref[eq6]{{{{}}}{(6)}},
represented by the simultaneous measurement
${\mathsf{M}}_{{{{\cal A}}}} (
{\text{\large $\times$}}_{k=1}^K {\mathsf{O}_k}$,
$ S_{[\rho]})$.

\par
\vskip0.3cm
\par
\par
\noindent
{\it Remark 2.}
[The relation between {{{}}}{\hyperlink{J}{(J)}} and {{{}}}{\hyperlink{K}{(K)}}]
Consider the {{{}}}{\hyperlink{K}{(K)}} in the classical cases,
i.e.,
${\cal A}=C_0(\Omega)$.
And assume the simultaneous observable
$F=F_1 {\text{\large $\times$}} F_2$ in
\hyperref[eq5]{(5)}.
Then,
putting $\rho_0=\delta_{\omega_0}$
(i.e., the point measure at $\omega_0$),
we see that
\begin{align*}
\hyperref[eq5]{{{{}}}{(5)}}=&
\frac{[F_1(\Xi_1) \times F_2(\Xi_2)](\omega_0)}{[F_1(\Xi_1) \times F_2(X_2)](\omega_0)}=[F_2(\Xi_2)](\omega_0)
\\
=&
\rho_0 (F_2(\Xi_2)).
\end{align*}
Since this equality holds for any ${\mathsf O}_2 =
(X_2, {\cal F}_{2}, F_2)$ and any $\Xi_2 \in{\cal F}_2$,
some may want to regard the $\rho_0$ as the state after the measurement
${\mathsf{M}}_{{{C_0(\Omega)}}} ({\mathsf{O}}_1{\; :=}(X_1, {\cal F}_1, {F}_1),S_{[\ast]})$
in the {{{}}}{\hyperlink{J}{(J)}}.
Thus, 
in spite of the {{{}}}{(\hyperlink{H2}{H$_2$})},
the {{{}}}{\hyperlink{J}{(J)}} may be
allowed in classical cases
if the $\rho_0$ may be regarded as something
represented by
the term
such as
{\lq\lq}imaginary 
state".
This is the meaning of
the informal {{{}}}{\hyperlink{J}{(J)}}.

\noindent
\begin{itemize}
\item[{\bf 4.4.}]
\bf
Sequential causal observable
and its realization
\end{itemize}

\par
Consider a tree
$(T{\; \equiv}\{t_0, t_1, \ldots, t_n \},$
$ \le )$
with the root $t_0$.
This is also characterized by
the map
$\pi: T \setminus \{t_0\} \to T$
such that
$\pi( t)= \max \{ s \in T \;|\; s < t \}$.
Let
$\{ \Phi_{t, t'} : {\cal A}_{t'}  \to {\cal A}_{t}  \}_{ (t,t')\in
T_\le^2}$
be a causal relation,
which is also represented by
$\{ \Phi_{\pi(t), t} : {\cal A}_{t}  \to {\cal A}_{\pi(t)}  \}_{ 
t \in T \setminus \{t_0\}}$.
Let an observable
${\mathsf O}_t{\; \equiv}
(X_t, {\cal F}_{t}, F_t)$ in the ${\cal A}_t$ 
be given for each $t \in T$.
Note that
$\Phi_{\pi(t), t}
{\mathsf O}_t$
$(
{\; \equiv}
(X_t, {\cal F}_{t},
\Phi_{\pi(t), t} F_t)$
)
is an observable in the ${\cal A}_{\pi(t)}$.

The pair
$[{\mathbb O}_T]
$
$=$
$[
\{{\mathsf O}_t \}_{t \in T}$,
$\{ \Phi_{t, t'} : {\cal A}_{t'}  \to {\cal A}_{t}  \}_{ (t,t')\in
T_\le^2}$
$]$
is called a
{\it sequential causal observable}.
For each $s \in T$,
put $T_s =\{ t \in T \;|\; t \ge s\}$.
And define the observable
${\widehat{\mathsf O}}_s
\equiv ({\text{\large $\times$}}_{t \in T_s}X_t, \boxtimes_{t \in T_s}{\cal F}_t, {\widehat{F}}_s)$
in ${\cal A}_s$
as follows:
\par
\noindent
\begin{align}
\widehat{\mathsf O}_s
&=
\left\{\begin{array}{ll}
{\mathsf O}_s
\quad
&
\!\!\!\!\!\!\!\!\!\!\!\!\!\!\!\!\!\!
\text{(if $s \in T \setminus \pi (T) \;${})}
\\
{\mathsf O}_s
{\text{\large $\times$}}
({}\bigtimes_{t \in \pi^{-1} ({}\{ s \}{})} \Phi_{ \pi(t), t} \widehat {\mathsf O}_t{})
\quad
&
\!\!\!\!\!\!
\text{(if $ s \in \pi (T) ${})}
\end{array}\right.
\label{eq7}
\end{align}
if
the commutativity condition holds
(i.e.,
if the product observable
${\mathsf O}_s
{\text{\large $\times$}}
({}\bigtimes_{t \in \pi^{-1} ({}\{ s \}{})} \Phi_{ \pi(t), t}
$
$\widehat {\mathsf O}_t{})$
exists)
for each $s \in \pi(T)$.
Using \hyperref[eq7]{(7)} iteratively,
we can finally obtain the observable
$\widehat{\mathsf O}_{t_0}$
in ${\cal A}_{t_0}$.
The
$\widehat{\mathsf O}_{t_0}$
is called the realization
(or,
realized causal observable)
of
$[{\mathbb O}_T]$.

\par
\par
\par
\noindent
\noindent

\par
\noindent
\vskip0.3cm
\par
\par
\noindent
{\it Remark 3.}
[Kolmogorov extension theorem]
In the general cases such that
countable additivity and infinite $T$
are required,
the existence of the above ${\widehat{\mathsf O}}_{t_0}$
is,
by using
the Kolmogorov extension theorem in probability theory
{{{}}}{\cite{Kolm}}, 
proved 
in the $W^*$-algebraic formulation
({\it cf.} {{{}}}{\cite{Ishi8}}).
We think that this fact is evidence that
the interpretation
{{{}}}{(\hyperlink{H2}{H$_2$})}
is hidden behind
the utility of the Kolmogorov extension theorem.
Recall the following well-known statement
that always appears
in the beginning of probability theory:
\begin{itemize}
\item[{{{}}}{\hypertarget{L}{(L)}}]
Let $(X, {\cal F}, P)$
be a probability space.
Then,
the probability
that
an event
$\Xi (\in {\cal F} )$
occurs
is given by
$P(\Xi)$,
\end{itemize}
which,
as well as Axiom 1,
is a statement without reality.
We consider that
the Kolmogorov extension theorem
is regarded as one of the finest answers to the
problem:
{\lq\lq}How should the statement {{{}}}{\hyperlink{L}{(L)}} be used?{\rq\rq}.
That is,
in 
mathematical probability theory,
the answer is presented as the form of a basic theorem
$\;$
(i.e.,
the Kolmogorov 
extension theorem).
On the other hand,
in measurement theory,
the problem {{{}}}{\hyperlink{G}{(G)}} is answered by the interpretation {{{}}}{\hyperlink{H}{(H)}}.
\par

\par
\par
\par
\noindent
\vskip0.3cm
\par
\par
\noindent
{\it Remark 4.}
[Wavefunction collapse]
Again reconsider the {{{}}}{\hyperlink{K}{(K)}} in the simplest case that
$T=\{t_0,t_1\}, \pi(t_1)=t_0$.
Taking a measurement
${\mathsf{M}}_{{{\cal A}_{t_0} }} ({\mathsf{O}}_{t_0},$
$ S_{[\rho_0]})$, 
we know that the measured value belongs to
$\Xi_0$
$(\in {\cal F}_{t_0})$.
Then, it may be usual to consider that
a certain wavefunction collapse happens
by the measurement,
that is,
${\frak S}^p  ({}{\cal A}^*_{t_0})$
$
\ni 
\rho_0 \mapsto{} \rho_0^{{}_{\Xi_0}}
\in {\frak S}^p  ({}{\cal A}^*_{t_0})
$.
And continuously, we take a measurement
${\mathsf{M}}_{{{\cal A}_{t_0} }} (\Phi_{t_0,t_1}{\mathsf{O}}_{t_1},$
$ S_{[\rho_{{}_0}^{{}_{\! \Xi_0}}
]})$.
Here, 
the probability that a measured value belongs to
$\Xi_1$
$(\in {\cal F}_{t_1})$
is,
by Axiom 1, given by
$\rho_0^{{}_{\Xi_0}}(\Phi_{t_0,t_1}F_1(\Xi_1))$.
However, this
$\rho_0^{{}_{\Xi_0}}(\Phi_{t_0,t_1}F_1(\Xi_1))$
must be equal to 
the conditional probability 
$$
\frac{\rho_0({\widehat F}
(\Xi_0 \times \Xi_1 ))}{\rho_0({\widehat F}
(\Xi_0 \times X_1 ))}
$$
if the commutativity condition holds
(i.e.,
the simultaneous observable ${\widehat{\mathsf O}}_{t_0}=
{{\mathsf O}_{t_0} \times \Phi_{t_0,t_1}{\mathsf O}_{t_1}}
$
$=(X_{t_0}{\text{\large $\times$}}X_{t_1} , {\cal F}_{t_0}\boxtimes {\cal F}_{t_1}, 
{\widehat F}{\; :=} {F}_{t_0}
{\text{\large $\times$}}
\Phi_{t_0,t_1}{F}_{t_1})$
exists).
This implies that
it suffices to consider only the measurement ${\mathsf{M}}_{{{\cal A}_{t_0} }} ({\mathsf{O}}_{t_0} {\text{\large $\times$}} \Phi_{t_0,t_1}{\mathsf{O}}_{t_1} ,$
$ S_{[\rho_0]})$.
That is,
two
measurements
${\mathsf{M}}_{{{\cal A}_{t_0} }} ({\mathsf{O}}_{t_0},$
$ S_{[\rho_{{}_0}^{}
]})$
and
${\mathsf{M}}_{{{\cal A}_{t_0} }} (\Phi_{t_0,t_1}{\mathsf{O}}_{t_1},$
$ S_{[\rho_{{}_0}^{{}_{\! \Xi_0}}
]})$
are not needed.
Also, if
the commutativity condition is ignored in the argument of
the wavefunction collapse,
it is doubtful.
\par

\par
\par
\noindent
{
\begin{itemize}
\item[{\bf 4.5.}]
{
\bf
Bayes' method
in classical
SMT
}
\end{itemize}
}

\par
\noindent

\rm 
Let ${\mathsf O}_1 \equiv (X, {\cal F}, F)$ be an observable in a commutative $C^*$-algebra $C(\Omega)$.
And let
${\mathsf O}_2 \equiv (Y, {\cal G}, G)$ be any observable in $C(\Omega)$.
Consider the product observable
${\mathsf O}_1 \times {\mathsf O}_2 \equiv (X\times Y, {\cal F} 
\boxtimes
{\cal G}, F \times G)$ in $C(\Omega)$.

Assume that
we know that
the measured value $(x,y)$ obtained by a simultaneous measurement
${\mathsf M}_{C(\Omega)}(
{\mathsf O}_1 \times {\mathsf O}_2,
S_{[*]}
{(\{\nu_0\})}
)$
belongs to
$\Xi \times Y \;(\in {\cal F} \boxtimes {\cal G} )$.
Then,
by Axiom$^{\rm S}$ 1,
we can infer that
\begin{itemize}
\rm
\item[{{{}}}{\hypertarget{M1}{(M$_1$)}}]
the probability $
P_\Xi (G(\Gamma))
$
that
$y$ belongs to $\Gamma (\in {\cal G})$
is given by
$$
\!\!\!
P_\Xi (G(\Gamma))
= \frac{\int_{\Omega} [F(\Xi)G(\Gamma)](\omega)
\;
\nu_0 (d \omega) }{
\int_{\Omega} [F(\Xi)](\omega)
\;
\nu_0(d \omega )}
\;\;
(\forall \Gamma \in {\cal G}).
$$
\end{itemize}
Thus,
we can assert that:

\begin{itemize}
\rm
\item[{{{}}}{\hypertarget{M2}{(M$_2$)}}]
[Bayes' method,
{\it cf.$\;$}{{{}}}{\cite{Ishi4, Ishi7}}].
\it
When
we
know that
a measured value
obtained by
a
measurement
${\mathsf M}_{C(\Omega)}(
{\mathsf O}_1 \equiv (X, {\cal F}, F)
, S_{[*]}
{(\{ \nu_0\})}
)$
belongs to
$\Xi$,
there is a reason to
infer
that
the mixed state after the measurement
is
equal to
$\nu_0^a$
$( \in {\cal M}_{+1}^m (\Omega ))$,
where
$$
\nu_0^a( D)= \frac{\int_{D} [F(\Xi)](\omega)
\;
\nu_0 (d \omega) }{
\int_{\Omega} [F(\Xi)](\omega)
\;
\nu_0(d \omega )}
\quad
(\forall D \in {\cal B}_\Omega ).
$$
\end{itemize}
\par
\noindent
This (as well as {{{}}}{\hyperlink{J}{(J)}}) is, of course, informal.

\par
\vskip0.2cm
\par
\noindent
{
\begin{itemize}
\item[{\bf 4.6.}]
{
\bf
Heisenberg's uncertainty principle
}
\end{itemize}
}

\par
\rm
In this section,
we use the $W^*$-algebraic formulation,
that is,
we consider the basic structure
$[B_c(H), B(H)]_{B(H)}$
({\it cf.} Section 5.5 later).
Let $A_1$ and $A_2$ be self-adjoint operators on a Hilbert space
$H$.
Note that the spectral representation
${\mathsf O}_k =( {\mathbb R}, {\cal B}_{{\mathbb R}}, E_{A_k} )$
of $A_k$
is regarded as the observable in $B(H)$.
Thus, we have two measurements
${\mathsf M}_{B(H)}( {\mathsf O}_k, S_{[\rho_u]} )$
$(\rho_u = |u \rangle \langle u | \in {\frak S}^p(B_c(H)^* ), k=1,2)$.
However, since ${\mathsf O}_1$ and ${\mathsf O}_2$ do not generally commute,
Interpretation
{{{}}}{(\hyperlink{H2}{H$_2$})} says that
it is impossible to
measure ${\mathsf M}_{B(H)}( {\mathsf O}_1, S_{[\rho]} )$
and
${\mathsf M}_{B(H)}( {\mathsf O}_2, S_{[\rho]} )$
simultaneously.
Then we consider another Hilbert space $K$,
$\rho_s= |s \rangle \langle s | \in {\frak S}^p(B_c(K)^* )$,
self-adjoint operators ${\widehat A}_k$ on a tensor Hilbert space
$H \otimes K$ such that
\begin{align*}
&
\text{{{{}}}{\hypertarget{N1}{(N$_1$)}}}:
\langle v, A_k v \rangle_H = 
\langle v \otimes s , {\widehat A}_k v \otimes s \rangle_{H \otimes K}
\quad
(\forall v \in H )
\\
&
\text{{{{}}}{\hypertarget{N2}{(N$_2$)}}}:
\text{
${\widehat A}_1$
and
${\widehat A}_2$
commute
}
\end{align*}
\par
\noindent
The existence is assured
({\it cf.} Theorem 1 in \cite{Ishi1}).
Define the observable
${\mathsf {\widehat O}}_k =( {\mathbb R}, {\cal B}_{{\mathbb R}}, 
E_{{\widehat A}_k} )$
in $B(H \otimes K )$
by the
spectral representation
of ${\widehat A}_k$. 
By the commutativity {{{}}}{\hyperlink{N2}{(N$_2$)}},
we get the product observable
${\mathsf {\widehat O}}_1 $
$\times $
${\mathsf {\widehat O}}_2$,
which is called the
approximately simultaneous observable of
$A_1$ and $A_2$.
Thus we can take an approximately simultaneous measurement
${\mathsf M}_{B(H \otimes K )}( {\mathsf {\widehat O}}_1 \times {\mathsf {\widehat O}}_2, S_{[\rho_u \otimes \rho_s]} )$
of $A_1$ and $A_2$.
Putting$\Delta_k = || ( {\widehat A}_k - A_k \otimes I)(u \otimes s) ||_{H \otimes K}$,
we have
the following Heisenberg's uncertainty principle
({\it cf.} Theorem 2 in \cite{Ishi1}):
\begin{align}
\Delta_1 \cdot \Delta_2 \ge 
\frac{1}{2}
|
\langle A_1 u , A_2 u \rangle_H -\langle A_2 u , A_1 u \rangle_H
|
\;\;
(\forall u \in H )
\label{eq8}
\end{align}
This should not be confused with Robertson's uncertainty principle
\cite{Robe},
which says
\begin{align}
\sigma_1 \cdot \sigma_2 \ge 
\frac{1}{2}
|
\langle A_1 u , A_2 u \rangle_H -\langle A_2 u , A_1 u \rangle_H
|
\;\;
(\forall u \in H )
\label{eq9}
\end{align}
where
$\sigma_k$
is defined by
$
|| (A_k- \langle u,  A_k u \rangle )u||_H
$.

\par
\vskip0.3cm
\par
\par
\noindent
{\bf
\large
5. Traditional Philosophies
(\hyperlink{fig1}{\bf Figure 1})
}
\par
\vskip0.5cm
\par
\par
\rm
\noindent
\par
According to \cite{Ishi3},
we shall explain \hyperlink{fig1}{\bf Figure 1}:[\textcircled{\scriptsize 1}-\textcircled{\scriptsize 8}] from the measurement theoretical point of view
in what follows..
\par
\noindent
\par
\vskip0.5cm
\par
\par
\noindent
\bf
5.1. Realistic world view and Linguistic world view
\par
\par
\vskip0.2cm
\par
\par
\rm
\hyperlink{fig1}{\bf Figure 1}
says that
the realistic world view \textcircled{\scriptsize R}
and the linguistic world view \textcircled{\scriptsize L}
exist together
in science.
Some may ask:
\begin{itemize}
\item[{{{}}}{\hypertarget{O}{(O)}}]
Why is
the series
\textcircled{\scriptsize L}
(or, idealism originated by Plato)
underestimated
in science?
\end{itemize}
We think that the reason
is due to the fact
that
the \textcircled{\scriptsize L}
is lacking in the axiomatization
\textcircled{\scriptsize 6}
if we do not have measurement theory.
That is, we believe that
there is no scientific world view without
axiomatization.

%

%

\par
\noindent
\vskip0.5cm
\par
\noindent
\par
\noindent
\bf
5.2. Dualism
\par
\par
\noindent
\vskip0.2cm
\par
\rm
Interpretation {{{}}}{(\hyperlink{H1}{H$_1$})} says
{\lq\lq}Image \hyperlink{fig2}{\bf Figure 2}
whenever measurement theory is used{\rq\rq}.
\par
\noindent
\vskip0.5cm
\noindent
\unitlength=0.5mm
\begin{picture}(200,72)(15,0)
\put(-8,0)
{
\allinethickness{0.2mm}
\drawline[-40](80,0)(80,62)(30,62)(30,0)
\drawline[-40](130,0)(130,62)(175,62)(175,0)
\allinethickness{0.5mm}
\path(20,0)(175,0)
%
\put(14,-5){
\put(37,50){$\bullet$}
}
\put(50,25){\ellipse{17}{25}}
\put(50,44){\ellipse{10}{13}}
\put(0,44){\put(43,30){\sf \footnotesize{observer}}
\put(42,25){\scriptsize{(I(=mind))}}
}
\put(7,7){\path(46,27)(55,20)(58,20)}
\path(48,13)(47,0)(49,0)(50,13)
\path(51,13)(52,0)(54,0)(53,13)
\put(0,26){
\put(142,48){\sf \footnotesize system}
\put(143,43){\scriptsize (matter)}
}
\path(152,0)(152,20)(165,20)(150,50)(135,20)(148,20)(148,0)
\put(10,0){}
\allinethickness{0.2mm}
\put(0,-5){
\put(130,39){\vector(-1,0){60}}
\put(70,43){\vector(1,0){60}}
\put(92,56){\sf \scriptsize \fbox{observable}}
\put(58,50){\sf \scriptsize }
\put(57,53){\sf \scriptsize \fbox{\shortstack[l]{measured \\ value}}}
\put(80,44){\scriptsize \textcircled{\scriptsize a}interfere}
\put(80,33){\scriptsize \textcircled{\scriptsize b}perceive a reaction}
\put(130,56){\sf \scriptsize \fbox{state}}
}
}
\put(30,-15){\bf
\hypertarget{fig2}{\bf Figure 2}. 
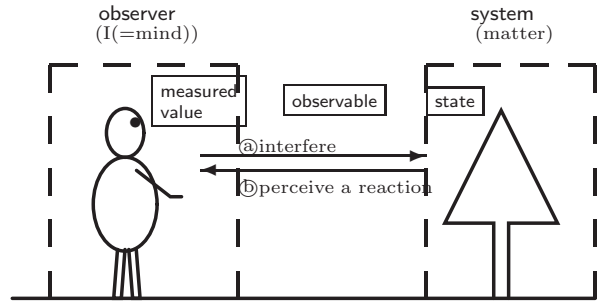
Descartes' figure
in MT
}
\end{picture}
\vskip1.3cm
\par
\noindent
where
the interaction
[\textcircled{\scriptsize a} and \textcircled{\scriptsize b}]
must not be emphasized, that is, it must be implicit.
That is because,
if it is explicitly stated,
the dualism
{{{}}}{(\hyperlink{H1}{H$_1$})} is violated.

John Locke's
famous
sayings
{\it
{\lq\lq}primary quality
{\rm
(e.g.,
length, weight, etc.)}{\rq\rq}}
and
{\it
{\lq\lq}secondary quality
{\rm (e.g.,
sweet, dark, cold, etc.)}{\rq\rq}}
urge us to associate the following correspondence:
$$
\text{state}\! \leftrightarrow \! \text{primary$\!$ quality},
\;
\text{observable} \! \leftrightarrow \! \text{second$\!$ quality}
$$
Also, it may be understandable to
regard {\lq\lq}observable{\rq\rq} as {\lq\lq}measurement instrument{\rq\rq}
or
{\lq\lq}sensory system{\rq\rq},
for example,
eyes,
glasses,
condensation trail,
etc.
And further,
it is natural to consider that
there is no measured value without observer's brain
(i.e.,
when something reaches observer's brain,
it becomes a measured value).
Thus, we want to consider the following correspondence
in {{{}}}{\hyperref[TB]{\bf Table 1}}.
\begin{table}[htbp] 
\small \caption{
Descartes vs. MT
\label{TB}}
\begin{center}
\begin{tabular}{
@{\vrule width 1.8pt\ }c
@{\vrule width 1.8pt\ }c|c|c
@{\vrule width 1.8pt }}
\noalign{\hrule height 1.8pt}
$\;\;$Descartes$\;\;$
&
$\;\;$mind (brain)$\;\;$
&
body
&
matter
\\
\noalign{\hrule height 1.8pt}
MT
& 
measured value
&
observable
& 
state
\\
\noalign{\hrule height 1.8pt}
\end{tabular}
\end{center}
\end{table}
\par
In the history of philosophy,
two kinds of dualisms (
based on
{\lq\lq}mind-body dualism{\rq\rq}
and
{\lq\lq}matter-mind dualism{\rq\rq}
)
may be frequently discussed.
However,
it should be noted that
the dualism
{{{}}}{(\hyperlink{H1}{H$_1$})}
is composed of
three concepts
as mentioned in {{{}}}{\hyperref[TB]{\bf Table 1}}.
Also,
the following question is nonsense
in the linguistic world view.
%
\begin{itemize}
\item[{{{}}}{\hypertarget{P1}{(P$_1$)}}]
What is {\lq\lq}measured value{\rq\rq} (
or, observable, state)?
$\;\;$
Or equivalently, in the sense of {{{}}}{\hyperref[TB]{\bf Table 1}},
what is {\lq\lq}mind{\rq\rq} (
or,
body, matter)?
$\;$
And moreover,
what is {\lq\lq}probability{\rq\rq}
(or,
causality)?
\end{itemize}
Therefore,
we must admit that
the correspondence in {{{}}}{\hyperref[TB]{\bf Table 1}}
is rather figurative,
that is,
it is not worth
discussing the problem {{{}}}{\hyperlink{P1}{(P$_1$)}} seriously
in the linguistic world view. 
From the great history of philosophy,
we learned that the serious consideration
(i.e.,
the consideration from the realistic world view) of the problem
{{{}}}{\hyperlink{P1}{(P$_1$)}}
(e.g.,
mind-body problem, etc.)
always led us into blind alleys.

We have to confirm that
we are now in the side of the linguistic world view
(i.e.,
after the linguistic turn
\textcircled{\scriptsize 3} in \hyperlink{fig1}{\bf Figure 1})
and not the realistic world view.
Thus, our interest always focuses on the problem:
\begin{itemize}
\item[{{{}}}{\hypertarget{P2}{(P$_2$)}}]
How should the term: {\lq\lq}measured value{\rq\rq} (or. observable, state
)
be used?
And moreover,
How should {\lq\lq}probability{\rq\rq}
(or,
causality) be described?
\end{itemize}
We,
of course,
assert that this answer is just given by measurement theory
(i.e.,
Axioms 1 and 2, Interpretation {{{}}}{\hyperlink{H}{(H)}}).
After accepting measurement theory,
what we can do in measurement theory is only to
trust in man's linguistic competence.
This is our linguistic world view {{{}}}{\hyperlink{I}{(I)}}.
\par
Here, we want to consider that
the following two are essentially the same:
\begin{itemize}
\item[{{{}}}{\hypertarget{Q1}{(Q$_1$)}}]
\it
To be is to be perceived
(by Berkeley)
\rm
\item[{{{}}}{\hypertarget{Q2}{(Q$_2$)}}]
\it
There is no science without measurement
(particularly,
measured value).
\end{itemize}
\rm
Also,
in the sense of {{{}}}{\hyperref[TB]{\bf Table 1}},
these are similar to
\begin{itemize}
\item[{{{}}}{\hypertarget{Q3}{(Q$_3$)}}]
\it
There is no science without human's brain,
\end{itemize}
\rm
which may be also similar to Kant's assertion
(see Section 5.7 later).
\par
\noindent
\vskip0.5cm
\par
\noindent
\par
\noindent
\bf
5.3.
I think, therefore I am.
\par
\rm
\par
\noindent
\vskip0.2cm
\par
\hyperlink{fig2}{\bf Figure 2} (Descartes' figure in MT) may be inspired from 
the Descartes primary principle:
\begin{itemize}
\item[{{{}}}{\hypertarget{R}{(R)}}]
\it
I think, therefore I am.
\end{itemize}
\rm
However, it should be noted that
the statement {{{}}}{\hyperlink{R}{(R)}} is not a statement in the dualism
of Interpretation {{{}}}{(\hyperlink{H1}{H$_1$})}.
That is because it is natural to
assume that
{\lq\lq}I{\rq\rq}={\lq\lq}observer{\rq\rq} and {\lq\lq}I{\rq\rq}={\lq\lq}system{\rq\rq}
in the {{{}}}{\hyperlink{R}{(R)}},
which clearly contradicts the {{{}}}{(\hyperlink{H1}{H$_1$})}.
We may see an irony
in the fact that
the non-dualistic statement
{{{}}}{\hyperlink{R}{(R)}} gives foundations to
the dualistic \hyperlink{fig2}{\bf Figure 2}.
However, it is sure that
the establishment of
{\lq\lq}I{\rq\rq}
in {{{}}}{\hyperlink{R}{(R)}}
brought us modern science
(\hyperlink{fig1}{\bf Figure 1}:\textcircled{\scriptsize 4}).

Also,
it is natural to
consider that
Heidegger's saying:
{\lq\lq}In-der-Welt-sein{\rq\rq}
is out of \hyperlink{fig2}{\bf Figure 2}, and thus, out of measure theory.
If some succeed the axiomatization of
{\lq\lq}In-der-Welt-sein{\rq\rq},
it will be the powerful rival of measurement theory
in science.

\par
\noindent
\vskip0.5cm
\par
\noindent
\par
\noindent
\bf
5.4. Causality and Probability
\rm
\vskip0.2cm
\par
The following paradigm shift
({\it cf.}
\hyperlink{fig1}{\bf Figure 1}:\textcircled{\scriptsize 1}
):
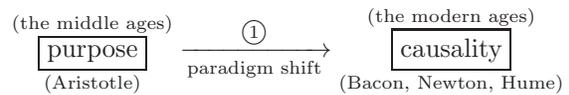
$$
\overset{\text{\scriptsize (the middle ages)}}{
\underset{\text{\scriptsize (Aristotle)}}{\fbox{\text{purpose}}}
}
\xrightarrow[\text{\scriptsize paradigm shift}]{{\textcircled{\scriptsize 1}}}
\overset{\text{\scriptsize (the modern ages)}}{
\underset{\text{\scriptsize (Bacon, Newton, Hume)}}{\fbox{\text{causality}}}
}
$$
is the greatest paradigm shift
throughout all history of science.
However,
it should be noted that
there are several ideas for
{\lq\lq}causality{\rq\rq}.
For example
Newton's causality is realistic,
and
Hume's causality is subjective.
On the other hand,
our causality (i.e.,
Axiom 2) is linguistic.
\par
Although 
some philosophers
(e.g.,
K. Popper
{{{}}}{\cite{Popp}})
consider that
the discovery of {\lq\lq}probability{\rq\rq}
is as great as
that of {\lq\lq}causality{\rq\rq},
it is sure that
the former is underestimated
in science,
The reason of the underestimation
may be due to
the fact
that
the {\lq\lq}probability{\rq\rq} is never presented in a certain world view
(but in mathematics ({\it cf.} {{{}}}{\cite{Kolm}}),
on the other hand,
the {\lq\lq}causality{\rq\rq} is established in the world view
(i.e.,
Newtonian mechanics).
We think that
it is desirable to
understand
the two concepts
(i.e.,
{\lq\lq}probability{\rq\rq} and {\lq\lq}causality{\rq\rq})
in the same world view.
It should be noted that this is realized
in Axioms 1 and 2
of
measurement theory.

\par
\noindent
\vskip0.5cm
\par
\noindent
\par
\noindent
\bf
5.5. Leibniz-Clarke Correspondence (Space-Time Problem)
\rm
\vskip0.2cm
\par
\noindent
\par
In this section,
first we must prepare the term
{\lq\lq}spectrum{\rq\rq}
in the formula \hyperref[eq3]{(3)}.
Consider the pair
$[{\cal A},\overline{\cal A}]_{B(H)}$,
called a
{\it
basic structure}
({\it cf.} Appendix in {{{}}}{\cite{Ishi2}}).
Here,
${\cal A} ( \subseteq B(H))$
is a $C^*$-algebra,
and
$\overline{\cal A}$
(${\cal A} \subseteq \overline{\cal A} \subseteq B(H)$)
is a particular $C^*$-algebra
(called a $W^*$-algebra)
such that
$\overline{\cal A}$ is the weak closure of
${\cal A}$
in $B(H)$.
Let
${\cal A}_S$
($\subseteq \overline{\cal A}$)
be the commutative $C^*$-subalgebra.
Note that
${\cal A}_S$
is represented
such that
${\cal A}_S$
$=C_0(\Omega_S)$
for some locally compact Hausdorff space $\Omega_S$
({\it cf.} {{{}}}{\cite{Murp}}).
The $\Omega_S$ is called a {\it spectrum}.
For example,
consider one particle quantum system,
formulated in a basic structure
$[B_c(L^2({\mathbb R}^3)),B(L^2({\mathbb R}^3))]_{B(L^2({\mathbb R}^3))}$.
Then, we can choose
the commutative $C^*$-algebra
$C_0({\mathbb R}^3)$
$( \subset
B(L^2({\mathbb R}^3)
)$,
and thus, we get the spectrum
${\mathbb R}^3$.
This simple example will make us proposal {{{}}}{\hyperlink{S2}{(S$_2$)}}
later.
\par
In
Leibniz-Clarke correspondence
(1715--1716),
they
(i.e.,
Leibniz
and Clarke(=Newton's friend)
)
discussed
{\lq\lq}space-time problem{\rq\rq}.
Their ideas are summarized as follows:
\begin{itemize}
\item[{{{}}}{\hypertarget{S1}{(S$_1$)}}]
$
\!\!
\cases
\!\!\!
\textcircled{\scriptsize R}
\!\!
:
\!\!
\underset{\text{\scriptsize (realistic world view)}}{\text{Newton, Clarke}}
&
\!\!\!\!\!\!\!\!
\cdots
\overset{\text{\scriptsize (space-time in physics)}}{
\underset{\text{\scriptsize }}{
\fbox{\text{realistic space-time}}}
}
\\
\\
\!\!\!
\textcircled{\scriptsize L}
\!\!
:
\!\!\!
\underset{\text{\scriptsize (linguistic world view)}}{\text{Leibniz}}  
&
\!\!\!\!\!\!\!\!
\cdots
\overset{\text{\scriptsize (space-time in language)}}{
\underset{\text{\scriptsize }}{
\fbox{\text{linguistic space-time}}}
}
\endcases
$
\end{itemize}
That is,
Newton considered
{\lq\lq}What is space-time?{\rq\rq}.
On the other hand,
Leibniz considered
{\lq\lq}How should space-time be represented?{\rq\rq},
though he did not propose
his language.
Measurement theory is in Leibniz's side,
and asserts that
\begin{itemize}
\item[{{{}}}{\hypertarget{S2}{(S$_2$)}}]
Space should be described as a kind of spectrum.
And
time should be described as a kind of tree.
In other words,
time is represented by a parameter
$t$
in a linear ordered tree $T$.
\end{itemize}
Therefore, we think that
the Leibniz-Clarke debates
should be essentially regarded as
{\lq\lq}the linguistic world view
\textcircled{\scriptsize L}{\rq\rq}
vs. {\lq\lq}the realistic world view \textcircled{\scriptsize R}{\rq\rq}.
Hence, the statement {{{}}}{\hyperlink{S2}{(S$_2$)}} should be added to
Interpretation
{{{}}}{\hyperlink{H}{(H)}} as sub-interpretation
of measurement theory.
\par
\noindent
\vskip0.5cm
\par
\noindent
\par
\noindent
\bf
5.6. Observer's time
\rm
\vskip0.2cm
\par
It is usual to consider that
quantum mechanics and observer's time are incompatible.
This leads 
Interpretation {{{}}}{(\hyperlink{H3}{H$_3$})},
which
says that
observer's time
is nonsense in measurement theory.
That is,
there is no tense
--- past, present, future ---
in measurement theory,
and therefore,
in science.
Many philosophers
(e.g.,
Augustinus,
Bergson,
Heidegger,
etc.)
tried to
understand
observer's time.
However,
from the scientific point of view,
their attempts may be reckless.
From the measurement theoretical point of view,
we feel sympathy
for
J. McTaggart,
whose paradox {{{}}}{\cite{McTa}} suggests that
observer's time 
leads science to inconsistency.

 \par
\noindent
\vskip0.5cm
\par
\noindent
\par
\noindent
\bf
5.7.
Linguistic turn
(\hyperlink{fig1}{\bf Figure 1}:\textcircled{\scriptsize 3},\textcircled{\scriptsize 5})
\rm
\vskip0.2cm
\par
For the question
{\lq\lq}Which came first, the world or the language?{\rq\rq},
two answers 
(the realistic world view
and the linguistic world view)
are possible.
However,
as mentioned in the {{{}}}{\hyperlink{I}{(I)}},
measurement theory is in the side of
{\lq\lq}the language came first{\rq\rq}
(due to
Saussure, Wittgenstein, etc,).

Note that two kinds of linguistic turns (i.e.,
\textcircled{\scriptsize 3} and \textcircled{\scriptsize 5})
are asserted in \hyperlink{fig1}{\bf Figure 1}.
That is,
reprinting the corresponding part of \hyperlink{fig1}{\bf Figure 1},
we see:
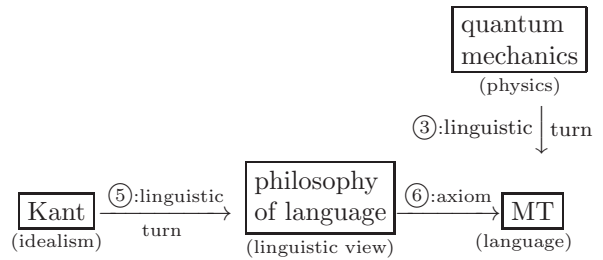
\begin{align*}
\CD
@. @. 
\!\!\!\!\!\!\!\!\!\!\!\!\!\!\!\!\!\!\!
\underset{\text{(physics)}}{\fbox{\shortstack[l]{quantum \\ mechanics}}}
\\
@. @. \!\!\!\!\!\!\!\!\!
@V{{\textcircled{\scriptsize 3}:\text{\footnotesize{linguistic}}}}V{\text{\footnotesize{turn}}\;\;}V
\\
\underset{\text{(idealism)}}{\fbox{\shortstack[l]{Kant}}}
\!\!\!\!
 @>{\textcircled{\scriptsize 5}:\text{linguistic}}>\text{turn }>
\underset{\text{(linguistic view)}}{
\fbox{\shortstack[l]{philosophy \\ of language}}
}
\!\!\!
@>{\textcircled{\scriptsize 6}:\text{axiom}}>>
\!\!\!\!\!\!\!\!\!\!\!\!\!
\underset{\text{$\;\;\;$(language)$\;\;\;$}\;\;\;}{\fbox{\shortstack[c]{MT}}}
\endCD
\end{align*}
where
\textcircled{\scriptsize 3}
is the turn from physics to language,
and
\textcircled{\scriptsize 5}
is the turn from
{\lq\lq}brain{\rq\rq}
to language.
Also, note that
the linguistic world view {{{}}}{\hyperlink{I}{(I)}} is essentially the
same as the following 
Wittgenstein's famous statement:
\begin{itemize}
\item[{{{}}}{\hypertarget{T1}{(T$_1$)}}]
\it
The limits of my language mean the limits of my world.
\end{itemize}
\rm
In fact,
Schr\"{o}dinger's cat
(or,
the theory of relativity)
is out of the world
described by measurement theory.
And moreover,
the assertion {{{}}}{\hyperlink{F}{(F)}} is similar to
Kant's main assertion
({\lq\lq}synthetic a priori judgement{\rq\rq})
in his famous book
{\lq\lq}Critique of Pure Reason {{{}}}{\cite{Kant}}{\rq\rq},
that is,
the two are similar
in the sense of
no experimental validation. 
Therefore, we want to consider the following correspondence:
\begin{itemize}
\item[{{{}}}{\hypertarget{T2}{(T$_2$)}}]
$
\underset{\text{(Axioms 1 and 2)}}{\text{[MT]}}
\leftrightarrow
\underset{\text{(synthetic a priori judgement
)}}{\text{
[Critique of Pure Reason]
}}
$
\end{itemize}
However,
as mentioned in
{{{}}}{\hyperlink{Q3}{(Q$_3$)}},
Kant's epistemological approach (i.e.,
Copernican turn)
is rather psychological
and not linguistic,
in spite that
his purpose is
philosophy
(i.e.,
the world view)
and not
psychology
(or, brain science).

We (as well as most scientists) consider:
\begin{itemize}
\item[{{{}}}{\hypertarget{T3}{(T$_3$)}}]
There is no boundary between
"mind" and "matter".
\end{itemize}
However, this fact should not be confused with
our assertion {{{}}}{\hyperlink{F}{(F)}},
that is,
\begin{itemize}
\item[{{{}}}{\hypertarget{T4}{(T$_4$)}}]
Most science should be described by
measurement theory
(i.e.,
the language based on dualism
(=
\hyperlink{fig1}{\bf Figure 1}
)
).
\end{itemize}
If there is the confusion between
{{{}}}{\hyperlink{T3}{(T$_3$)}}
and
{{{}}}{\hyperlink{T4}{(T$_4$)}},
this is due to the misreading of \cite{Kant}.
\par
\vskip0.5cm
\par
\noindent
\bf
5.8.
{\bf 
There are no facts, only interpretations
}
\rm
\vskip0.2cm
\par
\par
\noindent
\par
Nietzsche's famous saying
{\lq\lq}There are no facts, only interpretations"
is effective in measurement theory.
This is shown in this section.

\rm

\rm
Let testees drink water with various temperature
$\omega${$\;{}^\circ \!$C}$(0 {{\; \leqq \;}}\omega {{\; \leqq \;}}100)$.
And
you ask them {\lq\lq}cold"
or {\lq\lq}hot"
alternatively.
Gather the data,
(
for example,
$g_{c}(\omega)$ persons say {\lq\lq}cold",
$g_{h}(\omega)$ persons say {\lq\lq}hot")
and
normalize them,
that is,
get
the polygonal lines
such that
\begin{align*}
&
f_{c}(\omega)= \frac{g_{c}(\omega)}{\text{the numbers of testees}}
\\
&f_{h}(\omega)=\frac{g_{h}(\omega)}{\text{the numbers of testees}}
\end{align*}
And
\begin{itemize}
\item[]
$
f_{c} (\omega)
=
\cases
1 & \quad (0 {{\; \leqq \;}}\omega {{\; \leqq \;}}10 ) \\
\frac{70- \omega}{60}  & \quad (10 {{\; \leqq \;}}\omega {{\; \leqq \;}}70 ) \\
0 & \quad (70 {{\; \leqq \;}}\omega {{\; \leqq \;}}100 ) 
\endcases,
$
\\
\\
$
f_{h} (\omega) = 1- f_{c} (\omega)
$
\end{itemize}
\par
\par
\noindent
\noindent
Therefore, for example,
\begin{itemize}
\item[{{{}}}{\hypertarget{U1}{(U$_1$)}}]
You choose one person from the testees,
and
you ask him/her {\lq\lq}cold"
or {\lq\lq}hot"
alternatively.
Then the probability that
he/she
says
$
\left[\begin{array}{ll}
{}
\text{{\lq\lq}cold"}
%
{}
\text{{\lq\lq}hot"}
\end{array}\right]
$
is
given by
\rm
$
\left[\begin{array}{ll}
{}f_{\text \rm c}(55)=0.25
\\
{}
f_{\text \rm h}(55)=0.75
\end{array}\right]
$
\end{itemize}
This is described in terms of
Axiom 1
in what follows.
Consider
the state space
${\cal M}_{+1}^p (\Omega )$
$(\approx$
$\Omega$
$=$ interval $[0, 100](\subset {\mathbb R})$
)
and measured value space $X=\{c, h\}$.
Then,
we have
the (temperature) observable
${\mathsf O}_{ch}= (X , 2^X, F_{ch} )$
in 
$C ( \Omega )$
such that
\begin{align*}
&
[F_{ch}(\emptyset )](\omega ) = 0,
\quad
&{}&
[F_{ch}(X )](\omega ) = 1
\\
&
[F_{ch}(\{c\})](\omega ) = f_{c} (\omega ),
&{}&
[F_{ch}(\{h\})](\omega ) = f_{h} (\omega )
\end{align*}
Thus,
we get
a
measurement
${\mathsf M}_{C ( \Omega )} ( {\mathsf O}_{ch}, S_{[\delta_\omega]} )$.
Therefore,
for example,
putting
$\omega$=55{$\;{}^\circ \!$C},
we can,
by
{{{}}}{Axiom 1},
represent
the statement
{{{}}}{\hyperlink{U1}{(U$_1$)}}
as follows.
\begin{itemize}
\item[{{{}}}{\hypertarget{U2}{(U$_2$)}}]
the probability
that
a
{measured value}
$x(\in X {{=}}  \{c, h\})$
obtained by
{measurement}
\\
\\
${\mathsf M}_{C ( \Omega )} ( {\mathsf O}_{ch}, S_{[
\delta_{55}]} )$
belongs to
set
$
\left[\begin{array}{cc}
{}
\emptyset
\\
\{ \text{c}\}
\\
\{ {h} \}
\\
\{ {c} ,{h}\}
\end{array}\right]
$
is given by
\rm
$
\left[\begin{array}{ll}
{}
[F_{ch}( \emptyset )](55)= 0
\\
{}
[F_{ch}(
\{
{ c}
\}
)](55)= 
0.25
\\
{}
[F_{ch}(
\{
{ h}
\}
)](55)= 
0.75
\\
{}
[F_{ch}(
\{
{ c}
,
{ h}
\}
)](55)= 
1
\end{array}\right]
$
\end{itemize}

%

For example,
assume that
$\omega=5 \in \Omega (=[0,100])$.
Then, we see,
by {{{}}}{Axiom 1},
that
\begin{itemize}
\item[{{{}}}{\hypertarget{U3}{(U$_3$)}}]
the probability
that
a
{measured value}
$x(\in X {{=}}  \{c, h\})$
obtained by
the
{measurement}
${\mathsf M}_{C ( \Omega )} ( {\mathsf O}_{ch}, S_{[
\delta_5]} )$
belongs to
set
$
\left[\begin{array}{cc}
{}
\emptyset
\\
\{ \text{\it c}\}
\\
\{ \text{\it h} \}
\\
\{ \text{\it c} ,\text{\it h}\}
\end{array}\right]
$
is given by
\rm
$
\left[\begin{array}{ll}
{}
[F_{ch}( \emptyset )](5)= 0
\\
{}
[F_{ch}(
\{
{\text \rm c}
\}
)](5)= 
1
\\
{}
[F_{ch}(
\{
{\text \rm h}
\}
)](5)= 
0
\\
{}
[F_{ch}(
\{
{\text \rm c}
,
{\text \rm h}
\}
)](55)= 
1
\end{array}\right]
$
\end{itemize}
That is,
\begin{itemize}
\item[{{{}}}{\hypertarget{U4}{(U$_4$)}}]
a
{measured value}
$x(\in X {{=}}  \{c, h\})$
obtained by
{measurement}
${\mathsf M}_{C ( \Omega )} ( {\mathsf O}_{ch}, S_{[
\delta_5]} )$
is surely equal to
${\lq\lq}c"$.
\end{itemize}
Here,
we must not consider the following causality:
\begin{itemize}
\item[{{{}}}{\hypertarget{U5}{(U$_5$)}}]
$\qquad
\qquad
$
$
\underset{\text{\footnotesize {(}cause{)}}}{\fbox{\text{5{$\;{}^\circ \!$C}}}}
\xrightarrow[\text{(causality)}]{}
\underset{\text{\footnotesize (result)}}{\fbox{cold}}
$
\end{itemize}
The 
{{{}}}{\hyperlink{U4}{(U$_4$)}}
is not related to causality but measurement.
That is because
Axiom 2
(causality) is not used in {{{}}}{\hyperlink{U4}{(U$_4$)}}.
Recall that
\begin{itemize}
\item[{{{}}}{\hypertarget{U6}{(U$_6$)}}]
{{{}}}{Interpretation {{{}}}{(\hyperlink{H3}{H$_3$})}}
says that
causality belongs to the side of system
(and not between observer and system).
%
\end{itemize}
Thus, the {{{}}}{\hyperlink{U5}{(U$_5$)}} is not proper in the above situation.
\par
However, there is another idea as follows.
That is,
consider the dual causal operator
$
{\Phi^*}
:
{\cal M}_{+1}^p([0, 100])
\to
{\cal M}_{+1}^m (\{c, h\})$
such that
\begin{align*}
[{\Phi^*}
\delta_\omega
](D)
=
f_{ c }(\omega)
\cdot
\delta_{c}
(D)
+
f_{h }(\omega)
\cdot
\delta_{h}(D)
\\
\qquad
(\forall \omega \in [0,100] ,
\forall D \subseteq \{c, h\})
\end{align*}
Then,
the above {{{}}}{\hyperlink{U5}{(U$_5$)}}
can be regarded as the causality.
Considering
{the exact observable }${\mathsf O}^{\roman{(exa)}}$
(
=
$(\{c,h\}, 2^{\{c,h\}}, F_{\roman exa} )$
)
in
$C (\{c, h\})$
(where
$[F_{\roman exa}(\Xi)](\omega ) = 1 \; (\omega \in \Xi),$
$
=0  \; (\omega \notin \Xi) )$,
we see that
${\mathsf O}_{ch}=
\Phi {\mathsf O}^{\roman{(exa)}}$,
and thus,
we get the
{measurement}
${\mathsf M}_{C ( \Omega )} ( \Phi {\mathsf O}^{\roman{(exa)}},$
$S_{[
\delta_5]} )$.
In this case,
the above {{{}}}{\hyperlink{U5}{(U$_5$)}} is regarded as the causality.
In this sense, the {{{}}}{\hyperlink{U5}{(U$_5$)}} is proper.
\par
In the above argument,
readers may be reminded of
Nietzsche's famous saying:
\begin{itemize}
\item[{{{}}}{\hypertarget{U7}{(U$_7$)}}]
There are no facts, only interpretations.
\end{itemize}
Also,
note that
{\lq\lq}the measurement of a measurement"
is meaningless in measurement theory.

%
\par
\noindent
\vskip0.5cm
\par
\noindent
\par
\noindent
\bf
5.9.
Parmenides and Zeno
\rm
\vskip0.2cm
\par
About 2500 years ago,
Parmenides said that
\begin{itemize}
\item[{{{}}}{\hypertarget{V1}{(V$_1$)}}]
\it
There are no {\lq\lq}plurality{\rq\rq},
but only {\lq\lq}one{\rq\rq}. And therefore, there is no movement.
\rm
\end{itemize}
We want to regard
this {{{}}}{\hyperlink{V1}{(V$_1$)}}
as
the origin
of
Interpretation
{{{}}}{(\hyperlink{H2}{H$_2$})}
(i.e.,
{\lq\lq}Only one measurement is permitted.
And therefore,
a state never moves.{\rq\rq}).

\rm
\vskip0.2cm
\par
\noindent

Zeno, the student of Parmenides,
proposed several paradoxes
concerning
movement.
The following {\lq\lq}Achilles and the tortoise{\rq\rq}
is most famous.
\rm
\par
\noindent
{\rm [Achilles and the tortoise]}
\it
In a race, the quickest runner can never overtake the slowest, 
since the pursuer must first reach the point whence the pursued started, 
so that the slower must always hold a lead. 
\par
\rm

\par
Beginners of philosophy may have a question:
\begin{itemize}
\item[{{{}}}{\hypertarget{V2}{(V$_2$)}}]
Why have philosophers investigated such an easy problems during 2500
years?
\end{itemize}
However,
it should be noted that
{\lq\lq}Achilles and the tortoise{\rq\rq} is not
an elementary mathematical problem concerning
geometric series.
Since Parmenides and Zeno
were philosophers,
it is natural to consider that
Zeno's paradoxes should be regarded as
the problem concerning
world view.
That is,
we believe that
Zeno's question is as follows:
\begin{itemize}
\item[{{{}}}{\hypertarget{W}{(W)}}]
In what kind of world view
(in \hyperlink{fig1}{\bf Figure 1})
should Zeno's paradoxes be understood?
And further,
if the proper world view is not in \hyperlink{fig1}{\bf Figure 1},
propose the new world view
in which Zeno's paradoxes can be discussed!
\end{itemize}
It is clear that
Zeno's paradoxes are not in physics,
and thus,
Newtonian mechanics,
quantum mechanics
and the theory of relativity
and so on
are not proper
for the answer to the problem {{{}}}{\hyperlink{W}{(W)}}.
We assert that
classical measurement theory is
the proper world view,
in which Zeno's paradoxes are described.
The classical measurement theoretical description of
Zeno's paradoxes 
is easy,
in fact, it was simply shown in
{{{}}}{\cite{Ishi9}}.

Readers may be interested in
the unnatural situation
such that
{\lq\lq}Achilles{\rq\rq} and {\lq\lq}tortoise{\rq\rq}
are
quantum particles
(i.e.,
Zeno's paradoxes in quantum mechanics).
However we have no clear answer to this problem
though
our paper
{{{}}}{\cite{Inte}}
may be helpful.
Also,
for the more unnatural situation
(i.e.,
Zeno's paradoxes in the theory of relativity),
see {{{}}}{\cite{Zeno}}.

It is interesting and strange to see that
we already have the world description methods
(i.e.,
Newtonian mechanics,
quantum mechanics
and the theory of relativity)
for the unnatural situations,
but,
we have no world description method
for the natural situation
if we do not know measurement theory.
Thus,
if we believe in \hyperlink{fig1}{\bf Figure 1},
we are able to be convinced that
Zeno's paradoxes are solved.

\par
\noindent
\vskip0.5cm
\par
\noindent
\noindent
\bf
5.10.
Syllogism
\rm
\vskip0.2cm
\par
As an example of syllogism,
the following example
(due to Aristotle)
is famous.
\begin{itemize}
\item[{{{}}}{\hypertarget{X1}{(X$_1$)}}]
\it
Since Socrates is a man and all men are mortal,
it follows that Socrates is mortal.
\end{itemize}
\rm
However,
it should be noted that
there is a great gap
between
the {{{}}}{\hyperlink{X1}{(X$_1$)}} and the following mathematical syllogism:
\begin{itemize}
\item[{{{}}}{\hypertarget{X2}{(X$_2$)}}]
$A \Rightarrow B$,
$B \Rightarrow C$,
then,
it follows that
$A \Rightarrow C$.
\end{itemize}
\rm
That is because
the {{{}}}{\hyperlink{X2}{(X$_2$)}}
is merely mathematical rule
and not the world view,
and therefore,
it is not guaranteed that
the rule {{{}}}{\hyperlink{X2}{(X$_2$)}} is applicable to
the world {{{}}}{\hyperlink{X1}{(X$_1$)}}.
In fact,
as mentioned in ref. \cite{Ishi3},
the {{{}}}{\hyperlink{X1}{(X$_1$)}} does not hold in quantum cases.

Now, we are in the same situation such as {{{}}}{\hyperlink{W}{(W)}}.
That is, we have a similar question
(i.e.,
{{{}}}{\hyperlink{W}{(W)}}
$\approx
$
{{{}}}{\hyperlink{W'}{(W')}}
):
\begin{itemize}
\item[{{{}}}{\hypertarget{W'}{(W')}}]
In what kind of world view
(in \hyperlink{fig1}{\bf Figure 1})
should the phenomenon
{{{}}}{\hyperlink{X1}{(X$_1$)}}
be
described?
And further,
if the proper world view is not in \hyperlink{fig1}{\bf Figure 1},
propose the new world view!
\end{itemize}
We assert that
classical measurement theory is
the proper world view.
In fact,
in {{{}}}{\cite{Ishi5}},
the phenomenon {{{}}}{\hyperlink{X1}{(X$_1$)}}
is described as a theorem
in classical measurement theory.
\par
\noindent
\vskip0.5cm
\par
\noindent
\noindent
{\bf
\large
6. Heisenberg's Uncertainty Principle,
\\
$\quad$
EPR-Paradox and Syllogism}
\par
\vskip0.3cm
\noindent
\par
\noindent
\par
It is a matter of course that
any argument concerning
EPR-paradox is not clear
unless
the interpretation
is declared before the argument.
In this sense,
we believe that A. Einstein wanted to say in 
\cite{Eins}:
\begin{itemize}
\item[(Y)]
By a hint of EPR-paradox, propose a new quantum interpretation!
\end{itemize}
Now we have the linguistic interpretation of quantum mechanics,
and thus,
we can expect to clarify EPR-paradox.
\rm
\par
As shown and emphasized in {{{}}}{\cite{Ishi3}},
quantum syllogism does not generally hold.
We believe that
this fact was, for the first time, discovered in
EPR-paradox
{{{}}}{\cite{Eins}}.
The reason that we think so is as follows.

Consider the two-particles system composed of
particles $P_1$
and $P_2$,
which is formulated in
a Hilbert space
$L^2({\mathbb R}^2)$.
Let 
$\rho_{s} (\in {\frak S}^p (B_c( L^2({\mathbb R}^2)))$
be the EPR-state in EPR-paradox
(or, the singlet state in Bohm's situation).
Here, consider as follows:
\begin{itemize}
\item[{{{}}}{\hypertarget{Z1}{(Z$_1$)}}]
Assume that $(x_1,  p_2)$ and $p'_2$
are obtained by the simultaneous measurement
of [ the position of $P_1$, the momentum of $P_2$]
and
[the momentum of $P_2$].
Since it is clear that
$p_2=p'_2$, thus,
we see that
\begin{align*}
&
\underset{\text{\scriptsize
[the position of $P_1$, the momentum of $P_2$]
)}}{(x_1,  p_2)} \;\; 
\\
\\
&
\Longrightarrow 
\;\;
\\
&
\underset{\text{\scriptsize [the momentum of $P_2$]}}{p_2}
\end{align*}
\end{itemize}
Here, for the definition of {\lq\lq}$\Longrightarrow$",
see ref. {{{}}}{\cite{Ishi3}, or more precisely,
\cite{Ishi5}}.
\begin{itemize}
\item[{{{}}}{\hypertarget{Z2}{(Z$_2$)}}]
Assume that $p_1$ and $p_2$
are obtained by the simultaneous measurement
of [the momentum of $P_1$]
and
[the momentum of $P_2$].
Since 
the state
$\rho_{s} (\in {\frak S}^p (B_c( L^2({\mathbb R}^2)))$
is the EPR-state,
we see that $p_1=-p_2$,
that is, we see that
$$
\underset{\text{\scriptsize [
the momentum of $P_2$
]}}{p_2} \;\; \Longrightarrow 
\;\;
\underset{\text{\scriptsize [the momentum of $P_1$]}}{-p_2}
$$
\item[{{{}}}{\hypertarget{Z3}{(Z$_3$)}}]
Therefore,
if quantum syllogism holds,
{{{}}}{\hyperlink{Z1}{(Z$_1$)}}
and
{{{}}}{\hyperlink{Z2}{(Z$_2$)}}
imply that
\begin{align*}
\underset{\text{\scriptsize [the momentum of $P_1$]}}{- p_2}
\qquad
\end{align*}
that is, 
the momentum of $P_1$ is equal to $-p_2$.
\end{itemize}
Since the above
{{{}}}{\hyperlink{Z1}{(Z$_1$)}}-
{{{}}}{\hyperlink{Z3}{(Z$_3$)}}
is not the approximately simultaneous measurement
({\it cf.} the definition {\hyperlink{N2}{(N)}}),
it is not related to
Heisenberg's uncertainty principle
\hyperref[eq8]{{{{}}}{(8)}}.
Thus,
the conclusion {{{}}}{\hyperlink{Z3}{(Z$_3$)}}
is not contradictory to Heisenberg's uncertainty principle
\hyperref[eq8]{{{{}}}{(8)}}.
This was asserted in the remark 3 of
ref.
\cite{Ishi1}.
However,
now we can say that
the conclusion {{{}}}{\hyperlink{Z3}{(Z$_3$)}}
is not true.
That is because
%
the interpretation {{{}}}{(\hyperlink{H2}{H$_2$})}
(i.e., only one measurement is permitted)
says,
as seen in \cite{Ishi3},
that
quantum syllogism
does not hold by the non-commutativity of the above three
observables,
i.e.,
\begin{itemize}
\item[]
$
\cases
\text{[the position of $P_1$, the momentum of $P_2$]}
\\
\text{[the momentum of $P_2$]}
\\
\text{[the momentum of $P_1$]}
\endcases
$
\end{itemize}
Thus we see that
EPR-paradox is closely related to
the fact that
quantum syllogism does not hold in general.
This should be compared with Bell's inequality
\cite{Bell, Ishi2},
which is believed to be closely connected to
{\lq\lq}non-locality".

\par
\noindent
\vskip0.5cm
\par
\noindent
\noindent
{\bf
\large
7. Conclusions}
\par
\vskip0.3cm
\noindent
\par
As mentioned in
{{{}}}{\cite{Ishi2,Ishi3, Ishi4,Ishi5,Ishi6, Ishi7,Ishi8,
Ishi9 }},
measurement theory
( including several conventional system theories,
e.g.,
statistics, dynamical system theory, quantum system theory,
etc.)
is one of the most useful theories in science.
Following the well-known proverb:
{\it{\lq\lq}A sound mind in a sound body{\rq\rq}},
we consider:
{\it{\lq\lq}A good philosophy in a very useful theory.{\rq\rq}}
$\;$
For example,
{\lq\lq}A good realistic philosophy in the theory of relativity{\rq\rq}
is clearly sure.
Therefore, 
we
believe that
a good philosophy
has to be hidden behind measurement theory.
This belief makes us write this paper.
That is,
we want to assert
{\lq\lq}A good linguistic philosophy in quantum mechanics{\rq\rq}$\!\!.$

Dr. Hawking said in his best seller book {{{}}}{\cite{Hawk}}:
\it
Philosophers reduced the scope of their inquiries so much that 
Wittgenstein the most famous philosopher this century, 
said {\lq\lq}The sole remaining task for philosophy is 
the analysis of language.{\rq\rq} What a comedown 
from the great tradition of philosophy from Aristotle to Kant!
\rm
$\;\;$
We think that
this is not only his opinion but also most scientists' opinion.
And moreover, we mostly agree with him.
However,
we believe that
it is worth reconsidering
the series
\textcircled{\scriptsize L}
(i.e.,
the linguistic world view)
in \hyperlink{fig1}{\bf Figure 1}.
In spite of Lord Kelvin's saying:
\it
{\lq\lq}Mathematics is the only good metaphysics",
\rm
we assert that measurement theory is also good scientific metaphysics.
In order to answer the problem
{{{}}}{\hyperlink{B2}{(B$_2$)}}
(or, Zeno's paradoxes and so on),
we believe that
measurement theory is indispensable for science.

In this paper,
we see,
in the linguistic interpretation of quantum mechanics, that
EPR-paradox is closely related to
the fact that
quantum syllogism does not hold in general.
If we can believe that Bell's inequality is related to
{\lq\lq}non-locality",
we can clearly understand the difference between
EPR-paradox
and
Bell's inequality.

We hope that
our proposal will be examined from various view points.

\par
\par
\par
\noindent
\noindent

\rm
\par
\renewcommand{\refname}{
\large 
8. References}
{
\small

\normalsize
}


\begin{thebibliography}{9}
\rm
\bibitem{Ishi1} S. Ishikawa,
\newblock {\em Uncertainty relation
in simultaneous measurements for arbitrary observables,}
\newblock {\rm Rep. Math. Phys.,}
\newblock {\rm {\bf 9}},
\newblock {257-273},
\newblock {1991}
\\
\href{http://dx.doi.org/10.1016/0034-4877(91)90046-P}{doi: 10.1016/0034-4877(91)90046-P}
\bibitem{Robe} H.P. Robertson,
\newblock {\em The uncertainty principle,}
\newblock {\rm Phys. Rev.}
\newblock {{\bf 34}}, \newblock {163},
\newblock {1929}
\\
\href{http://dx.doi.org/10.1103/PhysRev.34.163}{doi:10.1103/PhysRev.34.163}
\bibitem{Eins}  A. Einstein, B. Podolsky and  N. Rosen,
\newblock {\em Can Quantum-mechanical Description of Physical Reality
be Considered Complete?}
\newblock {\rm Physical Review,}
{Vol. 47},
No. 10,
\newblock {{} 777-780},
1935
\\
\href{http://dx.doi.org/10.1103/PhysRev.47.777}{doi: 10.1103/PhysRev.47.777}
\bibitem{Ishi2} {S. Ishikawa,}
\newblock {\em A New Interpretation of Quantum Mechanics,}
\newblock {\rm Journal of quantum information science},
{Vol. 1}, No. 2, {}35-42,
2011
\\
\href{http://dx.doi.org/10.4236/jqis.2011.12005}{doi: 10.4236/jqis.2011.12005}
\bibitem{Ishi3} {S. Ishikawa,}
\newblock {\em Quantum Mechanics and the Philosophy of Language:
Reconsideration of traditional
philosophies,}
\newblock {\rm Journal of quantum information science},
{Vol. 2}, No. 1, {}2-9,
2012
\\
\href{http://dx.doi.org/ 10.4236/jqis.2012.21002}{doi: 10.4236/jqis.2012.21002}
\bibitem{Ishi4} {S. Ishikawa,}
\newblock {\em A Measurement Theoretical
Foundation of Statistics,}
\newblock {\rm Journal of Applied Mathematics},
{Vol. 3}, No. 3, {} 283-292,
2012
\\
\href{http://dx.doi.org/10.4236/am.2012.33044}{doi: 10.4236/am.2012.33044}
\bibitem{Ishi5}
S. Ishikawa,
\newblock {\em Fuzzy Inferences by Algebraic Method,}
\newblock {\rm Fuzzy Sets and Systems},
{Vol. 87}, No. 2, {}181-200,
1997
\\
\href{http://dx.doi.org/10.1016/S0165-0114(96)00035-8}{doi: 10.1016/S0165-0114(96)00035-8}
\bibitem{Ishi6}
S. Ishikawa,
{\em A Quantum Mechanical Approach to Fuzzy Theory,}
{{\rm Fuzzy Sets and Systems}}, 
{Vol. 90}, No. 3, {} 277-306,
1997
\\
\href{http://dx.doi.org/10.1016/S0165-0114(96)00114-5}{doi: 10.1016/S0165-0114(96)00114-5}
\bibitem{Ishi7}
S. Ishikawa,
{\it Statistics in measurements}, 
Fuzzy sets and systems, 
{Vol. 116}, No. 2, 141-154,
2000
\\
\href{http://dx.doi.org/10.1016/S0165-0114(98)00280-2}{doi:10.1016/S0165-0114(98)00280-2}
\bibitem{Ishi8}
S. Ishikawa,
{\em Mathematical Foundations of Measurement Theory,}
Keio University Press Inc. 335pages,
2006.
\\
(\url{http://www.keio-up.co.jp/kup/mfomt/})
\bibitem{Ishi9}
S. Ishikawa,
{\em Dynamical Systems, Measurements, Quantitative Language and
Zeno's Paradoxes,}
{\rm Far East Journal of Dynamical Systems,}
{Vol. 10}, No. 3, {} 277-292,
2008
\rm
\bibitem{Murp}
G. J. Murphy,
{\em $C^*$-algebras and Operator Theory,}
Academic Press, 
1990
\bibitem{Neum} {J. von Neumann,}
\newblock {\em Mathematical Foundations of Quantum Mechanics,}
\newblock {\rm Springer Verlag, Berlin,}
\newblock {1932}
\bibitem{Yosi}
K. Yosida,
{\em Functional Analysis,}
Springer-Verlag, 6th edition, 1980
\bibitem{Davi} E. B. Davies,
\newblock {\em Quantum Theory of Open Systems,}
\newblock {Academic Press,}
\newblock {1976}
\bibitem{Bohr} N. Bohr,
\newblock {\em Can Quantum-mechanical Description of Physical
Reality be Considered Complete?}
\newblock {\rm Physical Review,}
\newblock {{Vol. 47}, No. 8,{} 696-702}, 1935
\\
\href{http://dx.doi.org/10.1103/PhysRev.48.696}{doi:10.1103/PhysRev.48.696}
\bibitem{Kolm}
A. Kolmogorov,
{\em Foundations of the Theory of Probability (Translation),}
Chelsea Pub. Co. Second Edition,
New York,
1960,
\bibitem{McTa}
J. M. E. McTaggart,
{\em The Unreality of Time,}
{\rm
Mind
(A Quarterly Review of Psychology and Philosophy)}, Vol. 17,
{} 457-474, 1908
\bibitem{Popp}
K. R. Popper,
{\em The Logic of Scientific Discovery,}
Basic Books, Inc.
New York,
1959
\bibitem{Kant}
I. Kant,
{\em Critique of Pure Reason
(
Edited by P. Guyer,
A. W. Wood
),}
Cambridge University Press,
1999
\bibitem{Inte}
S. Ishikawa, T. Arai, T. Kawai,
{\em Numerical Analysis of Trajectories of 
a Quantum Particle in Two-slit Experiment,}
{\rm International Journal of Theoretical Physics,}
{Vol. 33}, No. 6,{} 1265-1274,
1993
\\
\href{http://dx.doi.org/10.1007/BF00670793}{doi: 10.1007/BF00670793}
\bibitem{Zeno}
J. Mazur,
{\em Motion Paradox,
The 2500-year-old Puzzle Behind all the Mysteries of
Time and Space,}
Dutton Adult,
Boston,
2007
\bibitem{Bell}
J. S. Bell, 
\newblock {\it On the Einstein-Podolsky-Rosen Paradox,}
\newblock {\rm Physics},
\newblock {Vol. 1}, 
\newblock {{} 195-200},
\newblock {1966}
\bibitem{Hawk}
S. Hawking,
{\em A Brief History of Time: From the Big Bang to Black Holes,
}
Bantam,
Boston,
1990
\end{thebibliography}
\end{document}